\documentclass[sigconf, authorversion]{acmart}

\usepackage{./Latex/macros}

\usepackage{array}
    \newcolumntype{P}[1]{>{\centering\arraybackslash}p{#1}}
\usepackage{booktabs}
\usepackage{caption}
\usepackage{float}
\usepackage{graphicx}
\usepackage{mathtools}
\usepackage{multirow}
\usepackage[normalem]{ulem}
\usepackage{placeins}
\usepackage{xcolor}

\AtBeginDocument{%
  \providecommand\BibTeX{{%
    \normalfont B\kern-0.5em{\scshape i\kern-0.25em b}\kern-0.8em\TeX}}}

\copyrightyear{2021} 
\acmYear{2021} 
\setcopyright{rightsretained} 
\acmConference[FAccT '21]{Conference on Fairness, Accountability, and Transparency}{March 3--10, 2021}{Virtual Event, Canada}
\acmBooktitle{Conference on Fairness, Accountability, and Transparency (FAccT '21), March 3--10, 2021, Virtual Event, Canada}
\acmDOI{10.1145/3442188.3445902}
\acmISBN{978-1-4503-8309-7/21/03}
\begin{document}

\title{Fairness in Risk Assessment Instruments: Post-Processing to Achieve Counterfactual Equalized Odds}

\author{Alan Mishler}
\affiliation{%
  \institution{Department of Statistics\\Carnegie Mellon University}
}
\orcid{0000-0002-7654-208X}

\author{Edward H. Kennedy}
\affiliation{%
  \institution{Department of Statistics\\Carnegie Mellon University}
}

\author{Alexandra Chouldechova}
\affiliation{%
  \institution{Heinz College\\Carnegie Mellon University}
}

\begin{abstract}
    In domains such as criminal justice, medicine, and social welfare, decision makers increasingly have access to algorithmic Risk Assessment Instruments (RAIs). RAIs estimate the risk of an adverse outcome such as recidivism or child neglect, potentially informing high-stakes decisions such as whether to release a defendant on bail or initiate a child welfare investigation. It is important to ensure that RAIs are fair, so that the benefits and harms of such decisions are equitably distributed.
    
    The most widely used algorithmic fairness criteria are formulated with respect to observable outcomes, such as whether a person actually recidivates, but these criteria are misleading when applied to RAIs. Since RAIs are intended to inform interventions that can reduce risk, the prediction itself affects the downstream outcome. Recent work has argued that fairness criteria for RAIs should instead utilize potential outcomes, i.e. the outcomes that \emph{would} occur in the absence of an appropriate intervention \cite{coston_counterfactual_2020}. However, no methods currently exist to satisfy such fairness criteria. 
    
    In this paper, we target one such criterion, \emph{counterfactual equalized odds}. We develop a post-processed predictor that is estimated via doubly robust estimators, extending and adapting previous post-processing approaches \cite{hardt_equality_2016} to the counterfactual setting. We also provide doubly robust estimators of the risk and fairness properties of  arbitrary fixed post-processed predictors. Our predictor converges to an optimal fair predictor at fast rates. We illustrate properties of our method and show that it performs well on both simulated and real data.
    
    \emph{Note: This is an extended version of the paper published in the FAccT '21 proceedings, with some additional theorems, proofs, and simulation results.}
\end{abstract}



\maketitle

\section{Introduction}
Machine learning is increasingly involved in high stakes decisions in domains such as healthcare, criminal justice, and consumer finance. In these settings, ML models often take the form of Risk Assessment Instruments (RAIs): given covariates such as demographic information and an individual's medical/criminal/financial history, the model predicts the likelihood of an adverse outcome, such as a dangerous medical event, recidivism, or default on a loan. Rather than rendering an automatic decision, the model produces a ``risk score,'' which a decision maker may take into account when deciding whether to prescribe a medical treatment, release a defendant on bail, or issue a personal loan.

The proliferation of machine learning has raised concerns that learned models may be discriminatory with respect to sensitive features like race, sex, age, and socioeconomic status. For example, there has been vigorous debate about whether a widely used recidivism prediction tool called COMPAS is biased against black defendants \citep{Angwin2016a, Angwin2016, Dieterich2016, Larson2016, Lowenkamp2016}. Concerns have also been raised about risk assessments used to identify high risk medical patients \citep{obermeyer_dissecting_2019} and about common credit scoring algorithms such as FICO \citep{rice_discriminatory_2012}, among many others. Collectively, these types of algorithms directly impact a large and growing swath of the global population.

These concerns have led to an explosion of methods in recent years for developing fair models and auditing the fairness of existing models. The most widely discussed fairness criteria impose constraints on the joint distribution of a sensitive feature, an outcome, and a predictor. These ``observational'' fairness criteria are inappropriate for RAIs, however. RAIs are not concerned with the \emph{observable outcomes} in the training data (``Did patients of this type historically experience serious complications?''), which are themselves a product of historical treatment decisions. Rather, they are concerned with the \emph{potential outcomes} associated with available treatment decisions (``\emph{Would} patients of this type experience complications \emph{if not treated?}''). Because treatments are not assigned at random---doctors naturally treat the patients they think are at high risk---these are distinct questions.

\citet{coston_counterfactual_2020} showed how RAIs that are optimized to predict observable rather than potential outcomes systematically underestimate risk for units that have historically been receptive to treatment, leading to suboptimal treatment decisions. They further showed how evaluations of the performance and fairness properties of RAIs with respect to observable outcomes are misleading. They proposed that RAIs should instead target counterfactual versions of standard performance and fairness metrics. However, they left open the question of how to develop predictors that satisfy such fairness notions.

In this paper, we develop a method to generate predictors that satisfy the fairness criterion \emph{approximate counterfactual equalized odds}. While many existing methods target observational fairness criteria \citep{kamiran_data_2012, hardt_equality_2016, Calmon2017, zafar_fairness_2017, donini_empirical_2018, narasimhan_learning_2018, kim_multiaccuracy_2019} and various types of causally motivated fairness \citep{Kilbertus2017, Kusner2017, nabi_fair_2018, nabi_learning_2019}, no methods currently exist that target counterfactual versions of standard observable fairness criteria like equalized odds. Our method post-processes an arbitrary existing predictor, extending previous post-processing methods \citep{hardt_equality_2016} to the counterfactual setting.

Our contributions are as follows. We first define approximate counterfactual equalized odds (\textsection \ref{section:notation}). After discussing related work (\textsection \ref{section:related_work}) and motivating the use of equalized odds over other candidate criteria (\textsection \ref{section:motivating_example}), we present a linear program that produces a loss-optimal post-processed predictor that satisfies this criterion (\textsection \ref{section:optimal_derived}). We provide theoretical results that our post-processed predictor is consistent in a particular sense at rates that depend on certain nuisance parameters. We show that our method performs well on both simulated and real data (\textsection \ref{section:results}).

\section{Notation and fairness definitions} \label{section:notation}
A table listing all notational choices can be found in Appendix \ref{appendix:notation}.

Let $A, D, Y$ denote a sensitive feature, decision, and outcome, respectively. We consider the setting in which all three are binary, though most of the definitions below extend readily to continuous settings. We define the counterfactual quantities of interest via the potential outcomes framework of \citep{neyman1923justification, Holland1986, rubin_causal_2005}. Denote by $Y^0, Y^1$ the potential (equivalently, ``counterfactual'') outcomes $Y^{D=0}, Y^{D=1}$. $Y_i^d$ is the outcome that would be observed for unit $i$ if, possibly contrary to fact, the decision were set to $D_i=d$. We refer to the two levels of the sensitive feature $A$ as the two ``groups,'' and we use ``treatment'' and ``intervention'' synonymously with ``decision.'' Let $S$ be any random variable that takes values in $\{0, 1\}$.

In most RAI settings, one of the decision options is a natural baseline corresponding to ``no intervention'' $(D = 0)$. Examples include the risk of recidivism if a defendant is released pretrial, or the risk of neglect or abuse if a child welfare call is not screened in for further investigation. Many or most RAIs do not generate a separate risk score for the outcome associated with intervention. In the case of child welfare, for example, call screeners must screen in any case in which a child is in apparent danger of neglect or abuse, regardless of the chances that a subsequent intervention will successfully prevent that neglect or abuse.

Denote the observational and counterfactual false positive rates of $S$ for group $a$ by $\FPR(S, a) = \Pb(S = 1 \mid Y = 0, A = a)$ and $\cFPR(S, a) = \Pb(S = 1 \mid Y^0 = 0, A = a)$. For example, $\cFPR(S,0)$ could represent the chance of being falsely labeled high-risk, among those black defendants who would not actually go on to recidivate if released pretrial, while $\cFPR(S, 1)$ could represent the corresponding error rate for white defendants who would not recidivate if released pretrial. Let $\FNR$, $\cFNR$, $\TPR$, and $\cTPR$ denote the corresponding observational and counterfactual false negative and true positive rates.
\begin{definition}
    A predictor $S$ satisfies \emph{observational equalized odds} (oEO) with respect to $A$ and $Y$ if $S \ind A  \mid  Y$. It satisfies \emph{counterfactual equalized odds} (cEO) if $S \ind A  \mid  Y^0$.
\end{definition}
When $A$, $Y$, and $S$ are all binary, equalized odds is equivalent to requiring that the corresponding false positive and false negative rates be equal for the two levels of $A$. Our post-processed predictor will target a relaxation of this criterion, defined below.
\begin{definition} \label{def:counterfactual_error_rate_diffs}
The \emph{counterfactual error rate differences} for a predictor $S$ are the differences $\Deltapos$ and $\Deltaneg$ in the $\cFPR$ and $\cFNR$ for the two groups $A = 0, A = 1$, defined as follows:
\begin{align*}
    \Deltapos(S) &= \cFPR(S, 0) - \cFPR(S, 1) \\
    \Deltaneg(S) &= \cFNR(S, 0) - \cFNR(S, 1) 
\end{align*}
\end{definition}
\begin{definition}
When $A, Y$, and $S$ are all binary, $\Rin$ satisfies \emph{approximate counterfactual equalized odds} with \emph{fairness constraints} $\epsilonpos, \epsilonneg \in [0, 1]$ if
\begin{align*}
    \left | \Deltapos(S)\right |  &\leq \epsilonpos \\
    \left | \Deltaneg(S)\right |  &\leq \epsilonneg
\end{align*}
\end{definition}
In general, a fairness-constrained predictor would not outperform an optimal unconstrained predictor, and in some cases, satisfying cEO exactly might degrade performance to the point that the RAI is no longer useful. This relaxation of cEO allows RAI designers to negotiate this tradeoff. This is similar in spirit to notions of approximate fairness that appear throughout the literature \citep{kearns_meritocratic_2017, donini_empirical_2018, menon_cost_2018}.

\section{Related work} \label{section:related_work}
\subsection{Observational and counterfactual fairness}
Equalized odds is one of several popular fairness criteria that impose constraints on the joint distribution of $(A, Y, S)$ \citep{barocas2018fairness}. These criteria appear under a variety of names. Equalized odds is known more generally as \emph{separation}, a term which covers settings in which these variables are not necessarily binary. The other two popular criteria in this class are \emph{independence} $(\Rin \ind A)$ and \emph{sufficiency} $(Y \ind A \mid \Rin)$. Independence also manifests as \emph{demographic parity}, \emph{statistical parity}, and \emph{group fairness}. Sufficiency is equivalent to \emph{calibration} or \emph{predictive parity} when all three variables are binary. Variants of all three criteria may be defined for example by conditioning on additional variables.

The counterfactual versions of these criteria simply replace $Y$ with the potential outcome $Y^0$ that is of interest \citep{coston_counterfactual_2020}. Note that these definitions cannot accommodate more than one potential outcome, such as the vector $(Y^0, Y^1)$, because only one of these outcomes is observed for each unit. This is the ``fundamental problem of causal inference'' \citep{Holland1986}.

Except in highly constrained, unrealistic conditions, these three criteria are pairwise unsatisfiable, regardless of whether they are defined with respect to $Y$ or $Y^0$ \citep{kleinberg2017, Chouldechova2017, barocas2018fairness}\footnote{See \citep{imai_principal_2020} for a set of sufficient conditions under which these unsatisfiability results disappear.}. We must therefore choose and justify which criterion we wish to target.

\subsection{Other causal fairness criteria}
The counterfactual fairness criteria just described consider potential outcomes with respect to a decision $D$. There is a distinct set of causally motivated fairness criteria that consider counterfactuals of the sensitive feature, or a proxy for the sensitive feature. They characterize a decision or prediction as fair if the sensitive feature or proxy does not ``cause'' the decision or prediction, either directly or along a prohibited pathway \citep{Kilbertus2017, Kusner2017, nabi_fair_2018, zhang_fairness_2018, nabi_learning_2019, wang_equal_2019}. There is some controversy over whether it is meaningful to discuss a counterfactual of a feature like race or gender \citep{VanderWeele2014, Glymour2014, hu_whats_2020}. Additionally, satisfying these metrics typically precludes use of most of the features that go into risk assessment, like prior history, which is not tenable in practice \citep{coston_counterfactual_2020}. Finally, it is not clear that counterfactuals of the sensitive feature are useful or appropriate to consider in the context of risk assessment. For example, in the child welfare setting, workers are compelled to screen in calls whenever a child is in danger of neglect or abuse. While it is important to ensure that risk is assessed accurately for different groups, it would be inappropriate to make screen-in decisions based on what a child's risk of neglect or abuse would be \emph{if they had been of a different race} their whole life, even if such an assessment were possible.

\subsection{Ways of achieving fairness} \label{subsection:ways_of_achieving_fairness}
There are three broad approaches to developing fair models: (1) preprocessing the input data to remove bias \citep{kamiran_data_2012, Calmon2017}, (2) constraining the learning process (aka ``in-processing'') \citep{zafar_fairness_2017, donini_empirical_2018, narasimhan_learning_2018}, and (3) post-processing a model to satisfy fairness constraints \citep{hardt_equality_2016, kim_multiaccuracy_2019}.

Our approach belongs to class (3). We refer to the predictor that our method returns equivalently as a ``post-processed'' or ``derived'' predictor. Each approach has advantages and disadvantages. Many widely used RAIs are proprietary tools developed by for-profit companies, so they are not amenable to internal tinkering. Developing new, fair(er) RAIs would be costly and perhaps infeasible from a policy perspective. The advantage of post-processing in this setting is that it can be applied to models that are already in use. The predictor that our method returns requires access at runtime only to the sensitive feature and the output of the existing predictor, so in principle, it could easily be incorporated into existing risk assessment pipelines.

In particular, our approach extends the work of \citet{hardt_equality_2016}, who proposed a method to post-process binary predictors to satisfy observational equalized odds (oEO) while minimizing loss with respect to observable $Y$. Their post-processed predictor is the solution to a simple linear program. We adapt their method to the counterfactual setting, in which the fairness criterion is approximate cEO and the loss function is weighted classification error with respect to $Y^0$. Because $Y^0$ is not observable when $D \neq 0$, we require tools from causal inference to solve this problem. Hardt et al.'s analysis treats the joint distribution of $(A, \Rin, Y)$ as known and frames post-processing primarily as an optimization problem. We build on their results by not making this assumption and treating post-processing as a statistical estimation problem.

\subsection{Why equalized odds?} \label{subsection:why_EO}
When evaluating a predictive system, it seems natural to focus on its real-world impact rather than its outputs per se. One desirable property of a decision process is the avoidance of disparate impact. Disparate impact is a legal doctrine enshrined in U.S. law that prohibits practices which have an unjustifiable adverse impact on people who share a protected characteristic, regardless of discriminatory intent. By way of shorthand, we will say that if $D \not\ind A  \mid  Y^0$, then the system exhibits discriminatory disparate impact\footnote{Some authors use ``disparate impact'' to refer to the criterion $S \ind A$, i.e. \emph{independence} \citep{zafar_fairness_2017}.}. In recidivism prediction, for example, this could mean that black defendants $(A = 0)$ who would not recidivate if released $(Y^0 = 0)$ are more likely to be detained pretrial $(D = 1)$ than white defendants who would not recidivate if released $(A = 1, Y^0 = 0)$. 

In the context of RAIs, decision makers typically have wide latitude in how they interpret and act on the risk scores, so constraining the RAI does not enforce fairness with respect to their decisions. However, if decision makers, after the introduction of the RAI, make their decisions only on the basis of the RAI scores and other variables $U$ which are independent of the RAI and $A$ given $Y^0$, then counterfactual equalized odds will imply $D \ind A  \mid  Y^0$. That is, let $D = f(S, U)$ represent the function $f$ describing the decision process after the RAI $S$ is introduced. If cEO is satisfied and $U \ind (S, A) \mid Y^0$, then it follows that $D \ind A \mid Y^0$. Even if $U \not\ind (S, A) \mid Y^0$, it is easy to see that if the conditional independence statement \emph{nearly} holds, or if $f$ depends primarily on $S$ rather than $U$, then discriminatory disparate impact can be small.

No such guarantees hold for predictors satisfying either independence or sufficiency. \citet{Chouldechova2017} in particular showed how predictors which satisfy sufficiency (predictive parity) are likely to yield decisions such that $D \not\ind A  \mid  Y$; these arguments are unchanged when we substitute $Y^0$ for $Y$. Though there is no consensus about how to quantify fairness, this is at least one consideration in favor of equalized odds.

\section{Motivating Example} \label{section:motivating_example}
Having motivated equalized odds over predictive parity or independence, we now motivate the use of counterfactual rather than observational equalized odds.

Consider a school district that assigns tutors to students who are believed to be at risk of academic failure. The school district wishes to develop a RAI, $S$, to better identify students who need tutors while ensuring that this resource is allocated fairly across two levels of the sensitive feature $A$. Let $D \in \{0, 1\}$ represent the decision to assign (1) or not assign (0) a tutor, and let $Y \in \{0, 1\}$ represent academic success (0) or failure (1).

A cEO predictor $S$ satisfies $\Pb(S \mid Y^0, A) = \Pb(S \mid Y^0)$, while an oEO predictor $S$ satisfies $\Pb(S \mid Y, A) = \Pb(S \mid Y)$. Divergence in these predictors is driven by the extent to which $Y \neq Y^0$ in the training data. In order to parameterize this divergence, we introduce the following definitions.
\begin{definition}
    The \emph{need rate} for group $a$ is $\Pb(Y^0 = 1 \mid A = a)$, the probability that a student from group $a$ would fail without a tutor.
\end{definition}
\begin{definition}
    The \emph{opportunity rate} for group $a$ is $\Pb(D = 1 \mid Y^0 = 1, A = a)$, the probability that a student in group $a$ who needs a tutor receives one.
\end{definition}
\begin{definition}
    The \emph{intervention strength} for group $a$ is $\Pb(Y^1 = 0 \mid Y^0 = 1, A = a)$, the probability that a student in group $a$ who would fail without a tutor would succeed with a tutor.
\end{definition}
We simulate a simple data generating process in which we allow the intervention strength to vary, while constraining it to be equal for the two groups. We fix all other parts of the distribution.  In particular, we set $\Pb(A = 1) = 0.7$, set the need rates to $0.4$ and $0.2$ for groups 0 and 1, and set the opportunity rates to $0.6$ and $0.4$. We set the probabilities that a tutor is assigned when it is not needed to $\Pb(D = 1 \mid Y^0 = 0, A = 0) = 0.3$ and $\Pb(D = 1 \mid Y^0  = 0, A = 1) = 0.2$. This represents a scenario in which the minority group has greater need, perhaps due to socioeconomic factors or prior educational opportunities, and also is likelier than the majority group to receive resources at baseline (prior to the development of the RAI). Finally, we set $\Pb(Y^1 = 0 \mid Y^0 = 0) = 1$, meaning that tutoring never \emph{increases} the risk of failure.

\begin{figure}
    \centering
    \includegraphics[width=\linewidth]{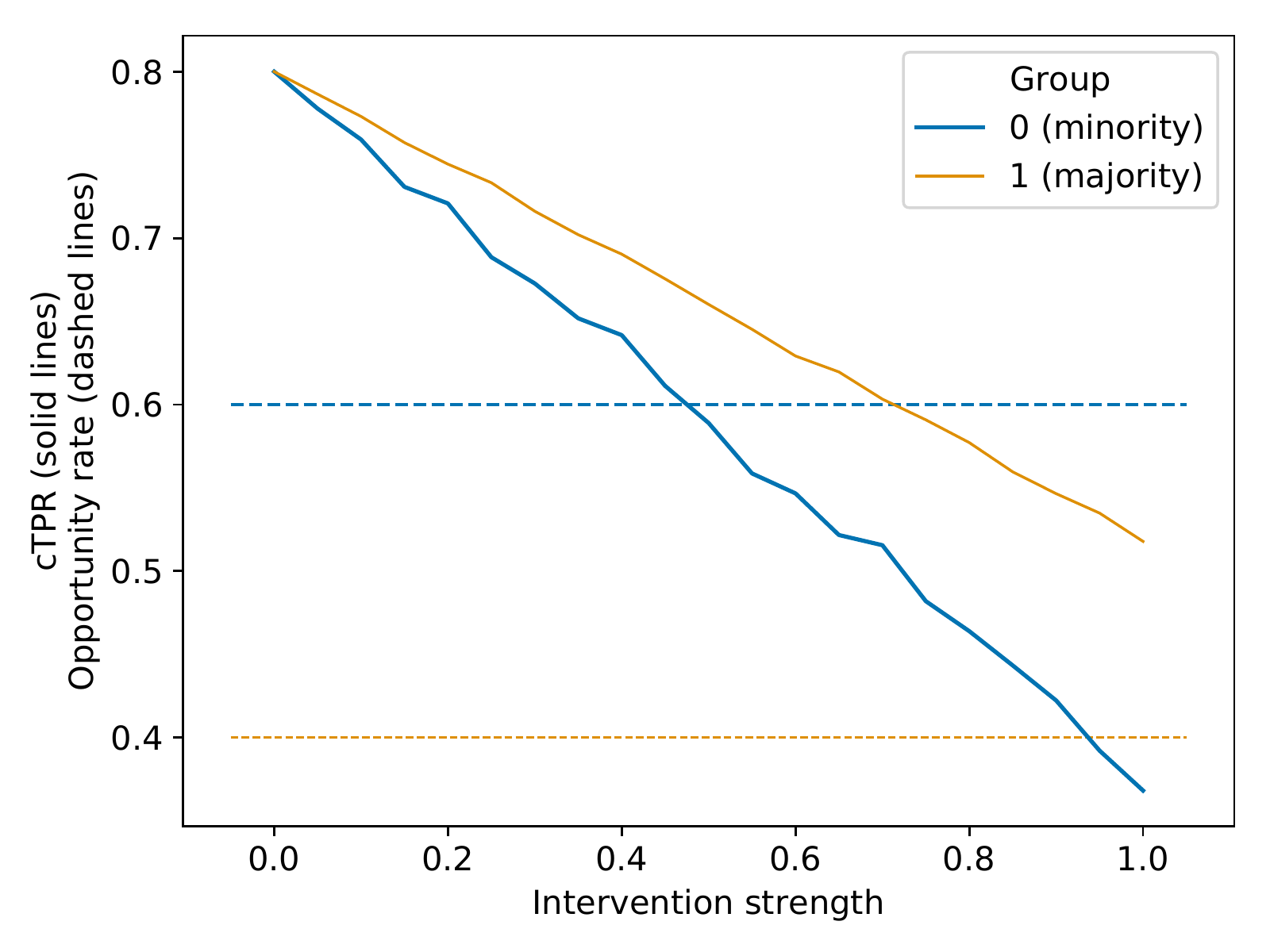}
    \caption{Counterfactual true positive rates (cTPRs; solid lines) for a RAI satisfying observational equalized odds (oEO), as a function of the intervention strength $\Pb(Y^1 = 0 \mid Y^0 = 1)$. Dashed lines indicate opportunity rates $\Pb(D = 1 \mid Y^0 = 0)$ prior to the development of the RAI. The more effective the tutoring (the higher the intervention strength), the worse the RAI is at identifying students who need it, and the greater the disparity in its performance between the minority and the majority group. When tutoring is more effective, the RAI may reduce the appropriate assignment of tutors below the baseline opportunity rates.}
    \label{fig:example}
\end{figure}

We consider a hypothetical oEO predictor $S$ with fixed false positive rate $\Pb(S = 1 \mid Y = 0, A) = \Pb(S = 1 \mid Y = 0) = 0.1$ and false negative rate $\Pb(S = 0 \mid Y = 1, A) = \Pb(S = 0 \mid Y) = 0.2$. We assume $S \ind Y^0 \mid A, Y$, as would be the case for example when $S$ is a high quality predictor of $Y$. Figure \ref{fig:example} shows the cTPRs (counterfactual true positive rates) for this predictor as a function of intervention strength, relative to the baseline opportunity rates for the two groups. When the intervention has no effect (strength 0), the cTPRs are equal because $Y \equiv Y^0$, so the cTPR and TPR are identical. (Of course, a strength of 0 means the tutoring is worthless.) For all strength values $> 0$, the cTPR of the minority group is lower than for the majority group. The difference in error rates increases as intervention strength increases. A cEO predictor avoids this problem by design: the cTPRs for the two groups are constrained to be equal.

This example makes it clear that oEO predictors in general will not prevent discriminatory disparate impact, whereas, as discussed in section \ref{subsection:why_EO}, counterfactual EO predictors have at least the potential to mitigate or avoid it.

This example also illustrates how oEO predictors can reduce rates of appropriate intervention. For example, suppose that decision makers, after the introduction of the RAI, set $D \equiv S$, i.e. they assign tutors precisely to students whom the RAI labels as high risk. Then, for any intervention strength $> 0.5$, the opportunity rate for the minority group decreases below baseline: the RAI \emph{harms the minority group}.

\section{An optimal fair derived predictor} \label{section:optimal_derived}

Having motivated counterfactual equalized odds, we now develop a method to generate predictors which satisfy it.
\subsection{Estimand} 
We expand our notation in order to fully describe our problem setting. Consider a random vector $Z = (A, X, D, \Rin, Y) \sim \Pb$, where in addition to the binary sensitive feature $A$, decision $D$, and outcome $Y$, we have covariates $X \in \R^p$ and a previously trained binary predictor $\Rin \in \{0, 1\}$. We require only that $\Rin$ is observable; we do not require access to its inputs or internal structure. $\Rin$ in practice could represent a RAI that is already in use, such as a recidivism prediction tool. The covariates $X$ may or may not overlap with the inputs to $\Rin$. Their role in the analysis is to render counterfactual quantities identifiable.

Our target is a derived predictor that satisfies approximate cEO. As in the case of observable equalized odds considered by \citet{hardt_equality_2016}, we achieve this by randomly flipping $\Rin$ with probabilities that depend only on $\Rin$ and $A$. Consider a column vector $\theta = (\theta_{0,0}, \theta_{0,1}, \theta_{1,0}, \theta_{1,1}) \in [0, 1]^4$. We define an associated derived predictor $\Rt$:
\begin{align*}
    &\Rt \sim \Bern(\theta_{A,S}) \\
    &\text{where } \theta_{A,S} = \sum_{a,\rin\in \{0, 1\}} \one\{A=a, \Rin=\rin\}\theta_{a, \rin}
\end{align*}
In other words, the $\theta_{a, 0}$ parameters represent conditional probabilities that $\Rin$ flips, while the $\theta_{a, 1}$ parameters represent conditional probabilities that $\Rin$ doesn't flip. Notice that for $\thetat = (0, 1, 0, 1)$, we have $\Rin_{\thetat} = \Rin$: the derived predictor is equal to the input predictor.

Our target is a loss-optimal fair predictor $\Ropt$, where the fairness criterion is approximate cEO. The loss function we consider is weighted classification error. For fixed $\theta$, denote the loss\footnote{We refer to this quantity as ``loss'' instead of the conventional ``risk'' in order to avoid confusion between risk assessment and the error rate of a predictor.} by $\risk(\Rt; \wfp, \wfn) = \wfp\Pb(Y^0 = 0, \Rt = 1) + \wfn\Pb(Y^0 = 1, \Rt = 0)$, where $\wfp, \wfn$ are chosen by the user to capture the relative importance of false positives and false negatives. (We will generally suppress the dependence of $\risk$ on $\wfp, \wfn$.) The estimand is
\begin{align*}
    \theta^* \in \argmin_{\theta} & \-\ \risk(\Rt) \\
    \text{subject to } & \-\ \theta \in [0, 1]^4 \\
                      & \-\ \left | \unfairFPR(\Rt)\right |  \leq \epsilonpos \\
                      & \-\ \left | \unfairFNR(\Rt)\right |  \leq \epsilonneg
\end{align*}
where $\Deltapos, \Deltaneg$ are given above in Definition \ref{def:counterfactual_error_rate_diffs}, and the fairness constraints $\epsilonpos, \epsilonneg \in [0, 1]$ are chosen by the user. Setting both these constraint parameters to 0 requires cEO to be satisfied exactly, while setting them to 1 allows $\Ropt$ to be arbitrarily unfair. Setting $\epsilonneg$ to 0 regardless of $\epsilonpos$ forces $\Ropt$ to satisfy \emph{counterfactual equal opportunity}; see \citet{hardt_equality_2016} for the observational definition of this criterion.

\begin{remark}
The full vector $Z$ is required only to estimate the parameter $\thetaopt$ that defines the optimal fair derived predictor. Once $\thetaopt$ has been estimated, the resulting derived predictor requires access at runtime only to the sensitive feature $A$ and the input predictor $\Rin$.
\end{remark}

Since our estimands involve counterfactual quantities, distributional assumptions are required in order to equate them to observable quantities.

\subsection{Identification} \label{section:identification}
In this subsection we show that the counterfactual error rates and loss can be identified under standard causal inference assumptions. All the quantities to be identified can be written in terms of the loss and the counterfactual error rates of the input predictor $\Rin$. For ease of notation, we first define two nuisance parameters that appear in the estimand and associated estimators, namely the outcome regression and propensity score function. We generally suppress the arguments of these functions in subsequent usage for the sake of conciseness. 
\begin{align*}
    \mu_0(A, X, \Rin) &= \E[Y \mid A, X, \Rin, D=0] \\
    \pi(A, X, \Rin) &= \Pb(D = 1 \mid A, X, \Rin)
\end{align*}
We make the following standard ``no unmeasured confounding''-type causal inference assumptions:
\begin{align*}
    & \text{A1. (Consistency) } \quad && Y = DY^1 + (1-D)Y^0 \\
    & \text{A2. (Positivity) } \quad && \exists \delta \in (0, 1) \text{ s.t. } \Pb(\pi(A, X, \Rin) \leq 1 - \delta) = 1 \\
    & \text{A3. (Ignorability) } \quad && Y^0 \ind D \mid A, X, S
\end{align*}
The consistency assumption means that the outcome observed for each individual is precisely the potential outcome corresponding to the treatment received. This implies that one person's treatment assignment does not affect another person's outcomes, meaning, for example, that an individual's recidivism behavior does not depend on whether other individuals are detained or released. The positivity or \emph{overlap} assumption requires that within strata of $(A, X, \Rin)$ of measure $> 0$, individuals have some chance of receiving no intervention. Finally, the ignorability or \emph{no unmeasured confounding} assumption requires that within strata of $(A, X, \Rin)$, the treatment $D$ is essentially random with respect to $Y^0$. Satisfying ignorability assumptions typically requires collecting a rich enough set of deconfounding covariates. In the present case, even if $X$ is low dimensional, the ignorability assumption is plausible if the input predictor $S$ substantially drives decision making, or if it happens to be an accurate (if not necessarily fair) predictor of $Y^0$.

Before giving the identifying expressions for $\risk(\Rt)$ and the error rate differences $\Deltapos, \Deltaneg$, we give identifying expressions for the error rates of the input predictor $\Rin$, which themselves appear in the expressions for $\Deltapos, \Deltaneg$.
\begin{proposition} \label{proposition:identification_input}
Under assumptions A1-A3, the counterfactual error rates of the input predictor $\Rin$ are identified as follows:
\begin{align*}
    \cFPR(\Rin, a) &= \frac{\E[\Rin(1 - \mu_0)\one\{A=a\}]}{\E[(1-\mu_0)\one\{A=a\}]} \\
    \cFNR(\Rin, a) &= \frac{\E[(1-\Rin)\mu_0\one\{A=a\}]}{\E[\mu_0\one\{A=a\}]}
\end{align*}
\end{proposition}
Proofs of propositions are given in Appendix \ref{appendix:propositions}. We now define several quantities that appear in the identifying expressions for $\risk(\Rt), \Deltapos$, and $\Deltaneg$:
\begin{align*}
    & \beta_{a, \rin} = \E\left[\one\{A=a,\Rin=\rin\}\left(\wfp - (\wfp + \wfn)\mu_0\right)\right], \text{ for } a, \rin \in \{0, 1\}  \\ 
    & \beta = (\beta_{0,0}, \beta_{0,1}, \beta_{1,0}, \beta_{1,1}) \\
    & \betapos = \left(1 - \cFPR(\Rin, 0), \-\ \cFPR(\Rin, 0), \-\ \cFPR(\Rin, 1) - 1,\-\  -\cFPR(\Rin, 1)\right) \\
    & \betaneg = \left(-\cFNR(\Rin, 0), \-\ \cFNR(\Rin, 0) - 1,\-\  \cFNR(\Rin, 1), \-\ 1 - \cFNR(\Rin, 1)\right)
\end{align*}

\begin{proposition} \label{proposition:identification_derived}
Under assumptions A1-A3, the loss and error rates of the derived predictor $\Rt$ are identified as:
\begin{align*}
    \risk(\Rt) &= \theta^T\beta + \wfn\E[\mu_0] \\
    \Deltapos(\Rt) &= \theta^T\betapos \\
    \Deltaneg(\Rt) &= \theta^T\betaneg
\end{align*}
\end{proposition}
Since the term $\wfn\E[\mu_0]$ in the loss is fixed, we can drop it without changing the minimizer of the loss. We can therefore rewrite the estimand as
\begin{align}
    \begin{split}
    \theta^* \in \argmin_{\theta} & \quad \theta^T\beta \label{eq:estimand_lp} \\
    \text{subject to } & \theta \in [0, 1]^4 \\
                      & | \theta^T\betapos |  \leq \epsilonpos \\
                      & | \theta^T\betaneg |  \leq \epsilonneg
    \end{split}
\end{align}
In other words, the optimal fair derived predictor is the solution to a linear program (LP). We refer to this as the ``true LP'' since it defines the estimand. We now define an estimator $\thetahat$ as the solution to an ``estimated LP.''

\subsection{Estimation} \label{subsection:estimation}

An estimator for $\thetaopt$ is derived by computing estimates $\betahat, \betaposhat, \betaneghat$ of the true LP coefficients and then solving the resulting estimated LP:
\begin{align*}
    \begin{split}
    \thetahat \in \argmin_\theta & \quad \theta^T\betahat \\
    \text{ subject to } & \quad \theta \in [0, 1]^4 \\
                        & \quad  | \theta^T\betaposhat |  \leq \epsilonpos \\
                        & \quad  | \theta^T\betaneghat |  \leq \epsilonneg
    \end{split}
\end{align*}

Any solution $\thetahat$ suffices. How should $\beta, \betapos, \betaneg$ be estimated? Before proposing a specific set of estimators, we first show that $\Rhat$ approaches optimal behavior at rates that depend on the performance of these estimators\footnote{We ignore optimization error, since this is a function of the number of optimization iterations and can be made arbitrarily small \citep{boyd_convex_2004}.}. We define two quantities of interest: the \emph{loss gap} and the \emph{excess unfairness}, and give accompanying theorems. Proofs of all theorems are given in Appendix \ref{appendix:theorems}.

Following standard usage, we say that an estimator $\psihat$ of a parameter $\psi$ is consistent at rate $f(n)$ for some real-valued function $f(n)$ if $\Vert \psihat - \psi \Vert = O_\Pb(1/f(n))$ for a suitable norm $\Vert \cdot \Vert$. For example, if $f(n) = \sqrt{n}$, then we say $\psihat$ converges at $\sqrt{n}$ rates. We say that an estimator converges \emph{faster} than $f(n)$ if $\Vert \psihat - \psi \Vert = o_\Pb(1/f(n))$. When $\psi \in \R^k$ for some $k$, the norm we are interested in is the Euclidean norm defined by $\Vert \psi \Vert^2 = \psi^T\psi$. When $\psi$ is a function of the random variable $Z$, the relevant norm is the $L_2$ norm with respect to $\Pb$, i.e., $\Vert \psi \Vert^2 = \int \psi(z)^2 d\Pb(z)$. For a random vector $\psi = (\psi_1, \ldots, \psi_k)$ taking values in $\R^k$, the norm is given by $\Vert \psi \Vert^2 = \sum_{j=1}^k \int \psi_j(z)^2 d\Pb(z)$.

\begin{definition}The \emph{loss gap} is $\risk(\Rhat) - \risk(\Ropt)$, the difference in loss between the derived predictor and the optimal derived predictor.
\end{definition} 
We use the term \emph{loss gap} rather than \emph{excess loss} to acknowledge that the loss of $\Rhat$ can be less than the loss of $\Ropt$, if $\thetahat$ falls outside the true constraints. Of course, this can only occur if $\Rhat$ violates the true fairness constraints, which can happen because the constraints are estimated.
\begin{theorem}[Loss gap]
Suppose that $\betahat, \betaposhat$, and $\betaneghat$ are all consistent at rate $f(n)$. Under Assumptions A1-A3:
\label{thm:risk_gap}
\begin{align*}
    \risk(\Rhat) - \risk(\Ropt) = O_\Pb(1/f(n))
\end{align*}
\end{theorem}
\begin{definition}
The \emph{excess unfairness of $\Rhat$ in the $\cFPR$} is 
\begin{align*}
    \UFP(\Rt) := \max\{ | \cFPR(\Rhat, 0) - \cFPR(\Rhat, 1) |  - \epsilonpos, 0\},
\end{align*}and the \emph{excess unfairness of $\Rhat$ in the $\cFNR$} is
\begin{align*}
    \UFN(\Rt) := \max\{ | \cFNR(\Rhat, 0) - \cFNR(\Rhat, 1) |  - \epsilonneg, 0\}
\end{align*}
\end{definition}
Since the estimated constraints should fluctuate around the true constraints, it's possible for $\Rhat$ to have error rate differences that are smaller than $\epsilonpos, \epsilonneg$, which motivates bounding these quantities below by 0.
\begin{theorem}[Excess unfairness]
Suppose that $\betahat, \betaposhat$, and $\betaneghat$ are all consistent at rate $f(n)$. Under assumptions A1-A3:
\label{thm:excess_unfairness}
    \begin{align*}
        \max\left\{\UFP(\Rhat), \UFN(\Rhat)\right\} = O_\Pb(1/f(n))
    \end{align*}
\end{theorem}
\begin{remark}\textbf{(The behavior of $\thetahat$ vs. $\Rhat$).} Without assumptions about how the loss and fairness of $\Rhat$ depend on $\thetahat$, there is no guarantee about the rate at which $\thetahat$ will approach $\thetaopt$. This is not a concern, however, since the object of interest is not $\thetaopt$ per se but a predictor that behaves like $\Ropt$.
\end{remark}

Arguably the simplest estimators for $\beta, \betapos, \betaneg$ involve plugging an estimate of the regression function $\mu_0$ into the identifying expressions in Proposition \ref{proposition:identification_input} and then computing empirical means in place of expectations. We propose instead using doubly robust, influence function-based estimators, which yield faster rates of convergence than plugin estimators in general nonparametric settings \citep{vaart_semiparametric_2002, tsiatis_semiparametric_2006}.

For ease of notation, let
\begin{align*}
     \phi &= \frac{1-D}{1-\pi}(Y - \mu_0) + \mu_0 
\end{align*}
denote the uncentered efficient influence function for $\E(Y^0)$, and let $\phihat$ denote an estimate constructed from estimates $\muhat_0$ and $\pihat$ \citep{bickel1993, Hahn1998, van_der_laan_unified_2003, kennedy_semiparametric_2016}. Both these nuisance functions can be estimated with arbitrary nonparametric learners. To minimize the use of indices, let $\Pn(f(Z)) = n^{-1}\sum_{i=1}^n f(Z_i)$ denote the sample average of any fixed function $f: \mathcal{Z} \mapsto \R$, so that for example $\Pn(\phihat)$ is the sample average of $\phihat(Z)$ after $\phihat$ has been constructed. The doubly robust estimators for individual coefficients are:
\begin{align}
    \betahat_{a,s} &= \Pn[\one\{A=a,S=s\}(\wfp - (\wfp + \wfn)\phihat)] \\
    \widehat{\cFPR}(\Rin, a) &= \frac{\Pn[\one\{A=a\}S(1-\phihat)]}{\Pn[\one\{A=a\}(1-\phihat)]} \label{eq:input_cFPR} \\
    \widehat{\cFNR}(\Rin, a) &= \frac{\Pn[\one\{A = a\}(1-S)\phihat]}{\Pn[\one\{A=a\}\phihat]} \label{eq:input_cFNR}
\end{align}
These estimates are assembled into the corresponding vectors $\betahat, \betaposhat, \betaneghat$.

In order to obtain optimal convergence rates, it is generally necessary to estimate the nuisance functions $\muhat_0$ and $\pihat$ on one sample and then compute the sample mean $\Pn$ on an independent sample conditional on those estimates. To obtain full sample size efficiency, one can swap the folds, repeat the procedure, and average the results, an approach that is popularly called cross-fitting \citep{bickel_estimating_1988, robins_higher_2008, zheng_asymptotic_2010, chernozhukov_doubledebiased_2018}. A $k$-fold version of cross-fitting with $k > 2$ is also possible. If $\muhat_0$ and $\pihat$ are assumed to be sufficiently ``well-behaved,'' i.e. if they belong to Donsker classes, then no such sample splitting is necessary. We prefer to avoid this assumption and utilize sample splitting. See Appendix \ref{appendix:sample_splitting} for a schematic of the sample splitting procedure.

The next theorem captures the double robustness property: under this sample splitting procedure, the coefficient estimators converge at a rate determined by the product of rates for the nuisance parameter estimators. Two additional mild assumptions are required.
\begin{align*}
    & \text{A4. (Bounded propensity estimator)} \hphantom{xxxxxxxxxxxxxxxxxxxx} \\
    & \quad \quad \exists \gamma \in (0, 1) \text{ s.t. } \Pb(\pihat(A, X, \Rin) \leq 1 - \gamma) = 1 \\
    & \text{A5. (Nuisance estimator consistency)} \\
    & \quad \quad \| \muhat_0 - \mu_0 \| = o_\Pb(1), \-\ \| \pihat - \pi \| = o_\Pb(1)
\end{align*}
Assumption A4 is the empirical analogue of the positivity assumption (A2). It can be trivially satisfied by truncating $\pihat$ at $1 - \delta$, the positivity threshold in assumption A2. Assumption A5 requires the nuisance parameter estimators to be consistent at any rate, which is reasonable if the nuisance parameters are estimated nonparametrically.

\begin{theorem}[Double robustness] \label{thm:double_robustness}
Suppose that $\Vert \muhat_0 - \mu_0 \Vert \Vert \pihat - \pi \Vert = O_\Pb(g(n))$ for some function $g(n)$. Under assumptions A1-A5:
\begin{align*}
    \Vert \betahat - \beta \Vert = O_\Pb\left(\max\left\{g(n), n^{-1/2}\right\}\right)
\end{align*}
and the same result holds for $\Vert \betaposhat - \betapos \Vert$ and $\Vert \betaneghat - \betaneg \Vert$.
\end{theorem}
\begin{corollary} \label{corollary:root_n}
If $\Vert \muhat_0 - \mu_0 \Vert \Vert \pihat - \pi \Vert = O_\Pb(n^{-1/2})$, then $\Vert \betahat - \beta \Vert = O_\Pb(n^{-1/2})$, and likewise for $\Vert \betaposhat - \betapos \Vert$ and $\Vert \betaneghat - \betaneg \Vert$.
\end{corollary}
The corollary shows that it is possible to obtain $\sqrt{n}$ convergence, the fastest rate attainable in general nonparametric settings, even when the nuisance parameters are estimated at slower than $\sqrt{n}$ rates. The condition of Corollary \ref{corollary:root_n} can be satisfied under relatively weak and nonparametric smoothness or sparsity assumptions \citep{gyorfi_distribution-free_2002, raskutti2011minimax}. For example, let $d = p + 2$ be the dimension of $(A, X, \Rin)$. If $\mu_0$ and $\pi$ are in H\"{o}lder classes with smoothness index $s > d/2$, then there exist nonparametric estimators $\muhat_0$ and $\pihat$ such that $\Vert \muhat_0 - \mu \Vert = o_\Pb(n^{-1/4})$ and $\Vert \pihat - \pi \Vert = o_\Pb(n^{-1/4})$, in which case the product of the rates would be $o_\Pb(n^{-1/2})$, which is faster than $O_\Pb(n^{-1/2})$.

By contrast, a plugin version of $\betahat$ would converge at a rate of $\sqrt{n}$ or the rate for $\Vert \muhat_0 - \mu_0 \Vert$, whichever is slower, and likewise for plugin versions of $\betaposhat$ and $\betaneghat$. Since $\sqrt{n}$ rates are generally unattainable in nonparametric regression, this means that plugin estimators would converge at slower than $\sqrt{n}$ rates, which, per Theorems \ref{thm:risk_gap} and \ref{thm:excess_unfairness}, would result in slower convergence in the loss gap and excess unfairness.

Corollary \ref{corollary:root_n} combined with Theorems \ref{thm:risk_gap} and \ref{thm:excess_unfairness} shows that if $\Vert \muhat_0 - \mu_0 \Vert \Vert \pihat - \pi \Vert = O_\Pb(n^{-1/2})$, then the excess risk and excess unfairness of $\Rhat$ decay at $\sqrt{n}$ rates.

\subsection{Estimating performance of the derived predictor}
Once $\thetahat$ has been computed, it is of interest to check both $\risk(\Rhat)$ and the error rate differences $\Deltapos(\Rhat), \-\ \Deltaneg(\Rhat)$ of the resulting derived predictor $\Rhat$, for example to understand the performance ``cost'' of fairness and to check whether the procedure successfully controlled the error rate differences.

These estimates should be computed on a test set that is independent of the sample used to estimate $\thetahat$. Within the test set, the same sample splitting considerations apply: unless they are assumed to belong to Donsker classes, the nuisance parameters $\muhat_0$ and $\pihat$ should be estimated on separate folds from the folds used to compute the relevant sample means $\Pn$.

Since the estimators below are conditional on a fixed $\thetahat$, they can in fact be applied to any fixed parameter value $\theta \in [0, 1]^4$. We define two additional quantities of interest.
\begin{definition}
    The \emph{loss change} for a derived predictor $\Rt$ relative to an input predictor $\Rin$ is $\riskchange(\Rt) = \risk(\Rt) - \risk(\Rin)$.
\end{definition}
We refer to a \emph{loss change} rather than an \emph{increase in loss} because it is possible for $\Rt$ to have smaller loss that $\Rin$. This is not a typical expectation: in fair prediction problems, the set of fair classifiers is necessarily smaller than the set of fair and unfair classifiers, so there is a fairness-accuracy tradeoff. In the RAI setting, however, since predictors are typically trained to predict observable outcomes, their performance may be arbitrarily bad with respect to the potential outcome $Y^0$. It is therefore not implausible than a derived fair predictor could have higher accuracy than the input predictor.

\begin{definition}
    The \emph{predictive change} for a derived predictor $\Rt$ relative to an input predictor $\Rin$ is $\Pb(\Rhat \neq \Rin)$.
\end{definition}
The predictive change is the proportion of input predictions that the post-processed predictor flips, which gives a measure of the effect of post-processing.

Once again, we propose using doubly robust estimators. These estimators are essentially identical to the estimators of the LP coefficients used in the previous section. Here, however, we are interested in properties of these estimators, rather than properties of our derived predictor $\Rhat$. In particular, we are interested in deriving confidence intervals, in addition to guaranteeing rates of convergence. The estimators are
\begin{align}
    & \riskhat(\Rt) = \theta^T\betahat + \wfn\Pn(\phihat) \tag{loss} \\
    & \riskchangehat(\Rt) = (\theta - \thetat)^T\betahat \tag{loss change} \\
    & \widehat{\cFPR}(\Rt, a) = \theta_{a,0}\left(1 - \widehat{\cFPR}(\Rin, a)\right) + \theta_{a,1}\widehat{\cFPR}(\Rin, a) \quad \tag{cFPR} \\
    & \widehat{\cFNR}(\Rt, a) = (1-\theta_{a,0})\left(\widehat{\cFNR}(\Rin, a)\right) + \theta_{a,1}\left(1 - \widehat{\cFNR}(\Rin, a)\right) \tag{cFNR} \\
    & \Deltahatpos(\Rt) = \theta^T\betahat^{+} \tag{error rate difference in cFPR} \\
    & \Deltahatneg(\Rt) = \theta^T\betahat^{-} \tag{error rates difference in cFNR}
\end{align}

where recall $\thetat = (0, 1, 0, 1)$, so that $\Rin_{\thetat} = \Rin$ (i.e., the derived predictor is simply the input predictor). Note that the loss estimator adds back in the portion of the loss that doesn't depend on $\theta$ and that we consequently removed from the LP in \eqref{eq:estimand_lp}.

The predictive change does not involve counterfactual quantities, so it can be straightforwardly estimated with a plugin estimator: $\widehat{\Pb}(\Rhat \neq \Rin) = \Pn\left\{\Pb\left(\Rhat \neq \Rin|A, \Rin\right)\right\} =$
\begin{align*}
     \Pn\left\{\sum_{a\in\{0,1\}}\left[\theta_{a,0}\one\{A = a, \Rin = 0\} + (1 - \theta_{a, 1})\one\{A = a, \Rin = 1\}\right]\right\}
\end{align*}
Since this is a sample average, it is asymptotically normal, and confidence intervals can be derived via the central limit theorem. In order to obtain asymptotic normality for the remaining estimators, we require the produce of nuisance parameter errors to decay faster than $\sqrt{n}$:
\begin{align*}
    & \text{A6. (Nuisance estimator rates).} \hphantom{xxxxxxxxxxxxxxxxxxxx} \\
    & \quad \quad  \Vert \muhat_0 - \mu_0 \Vert \Vert \pihat - \pi \Vert = o_\Pb(1/\sqrt{n})
\end{align*}
As described above, assumption A6 can be satisfied under sparsity or smoothness conditions on $\mu_0$ and $\pi$.

\begin{theorem}[Asymptotic normality] \label{thm:asymptotic_normality}
Fix $\theta \in [0, 1]^4$. Under assumptions A1-A6:
\begin{align*}
    & \sqrt{n}\left(\widehat{\risk}(\Rt) - \risk(\Rt)\right) \rightsquigarrow N\left(0, \var\left(f_\theta\right)\right) \\
    & \sqrt{n}\left(\riskchangehat - \riskchange\right) \rightsquigarrow N\left(0, \var\left(f_\theta - f_{\thetat}\right)\right) \\
    & \sqrt{n}\left(\widehat{\cFPR}(\Rt, a) - \cFPR(\Rt, a)\right) \rightsquigarrow N\left(0, \var\left(g_a\right)\right) \\
    & \sqrt{n}\left(\widehat{\cFNR}(\Rt, a) - \cFNR(\Rt, a)\right) \rightsquigarrow N\left(0, \var\left(h_a\right)\right) \\
    & \sqrt{n}\left(\Deltahatpos - \Deltapos\right) \rightsquigarrow N\left(0, \var(g_0 - g_1)\right) \\
    & \sqrt{n}\left(\Deltahatneg - \Deltaneg\right) \rightsquigarrow N\left(0, \var(h_0 - h_1)\right)
\end{align*}
where
\begin{spreadlines}{1em}
\begin{align*}
    f_\theta &= \theta_{A,\Rin}(\wfp - (\wfp + \wfn)\phi) + \wfn\phi \\
    g_a &= (\theta_{a,1} - \theta_{a,0})\frac{\one\{A=a\}(1 - \phi)(\Rin - \cFPR(\Rin, a))}{\E[\one\{A=a\}(1 - \phi)]} \\
    h_a &= (\theta_{a,1} - \theta_{a,0})\frac{\one\{A=a\}\phi(\Rin - \cFNR(\Rin, 0))}{\E[\one\{A=a\}\phi]}
\end{align*}
\end{spreadlines}
and recall
\begin{align*}
    \theta_{A,\Rin} &= \sum_{a,\rin\in{0,1}}\theta_{a,\rin}\one\{A=a, \Rin=\rin\}
\end{align*}
and the estimators $\riskhat, \-\ \riskchangehat, \widehat{\cFPR}, \-\ \widehat{\cFNR}, \-\ \Deltahatpos, \-\ \Deltahatneg$ attain the nonparametric efficiency bound, meaning that no other estimator has smaller asymptotic variance.
\end{theorem}
\begin{corollary} \label{corollary:risk}
Given a consistent estimator for $\var(f_\theta)$, an asymptotically valid 95\% confidence interval for $\risk(\Rt)$ is given by $\riskhat(\Rt) \pm 1.96\cdot\widehat{\var}(f_\theta)/\sqrt{n}$. An asymptotically valid test of the hypothesis $\risk(\Rt) = C$ for any $C$ consists of evaluating whether $C$ is in the confidence interval. Analogous results hold for $\riskchange, \-\ \cFPR, \cFNR, \Deltapos, \Deltapos$.
\end{corollary}

Perhaps the most natural estimators for these variances are the sample variances of $\fhat_\theta, \-\ \ghat_a, \-\ \hhat_a, \-\ \ghat_0 - \ghat_1$, and $\hhat_0 - \hhat_1$, where these quantities are defined by the following:
\begin{spreadlines}{1em}
\begin{align*}
    \fhat_\theta &= (\wfp - (\wfp + \wfn)\phihat)\theta_{A,\Rin} + \wfn\phihat \\
    \ghat_a &= (\theta_{a,1} - \theta_{a,0})\frac{\one\{A=a\}(1 - \phihat)(\Rin - \widehat{\cFPR}(\Rt, a))}{\Pn[\one\{A=a\}(1 - \phihat)]^{-1}} \\
    \hhat_a &= (\theta_{a,1} - \theta_{a,0})\frac{\one\{A=a\}\phihat(\Rin - \widehat{\cFNR}(\Rt, 0))}{\Pn[\one\{A=a\}\phihat]^{-1}}
\end{align*}
\end{spreadlines}
The quantities $f_\theta, g_a$, and $h_a$ are the efficient influence functions for the loss and error rates.

\section{Results} \label{section:results}
There is no previous method designed to achieve counterfactual equalized odds or related fairness criteria that we can compare our method to. We instead compare our method to an approach that uses plugin estimators for the LP coefficients, in order to illustrate the advantages of the doubly robust estimators.
\subsection{Simulations} \label{subsection:simulations}
We use one set of simulations to illustrate Theorems \ref{thm:risk_gap}-\ref{thm:double_robustness} and another set to explore fairness-performance tradeoffs. We use equal misclassification weights $\wfp = \wfn = 1$, so that false positives and false negatives contribute equally to the loss. Simulations illustrating Theorem \ref{thm:asymptotic_normality} can be found in Appendix \ref{appendix:asymptotic_normality}.

Each estimation procedure was run 500 times for each sample size $n \in \{100, 200, 500, 1000, 5000, 20000\}$. Since $\mu_0$ is known here, the ``true'' loss and fairness values were computed on a separate validation set of size 500,000, using plugin estimators with the true $\mu_0$. These values showed negligible variation over many repetitions.

\subsubsection{Setup}
First, we define a \emph{pre-RAI} data generating process. Using this data, we train a predictor $\Rin$ to predict observable outcomes $Y$, mirroring how RAIs are typically constructed in practice. We then define a \emph{post-RAI} data generating process, which only differs in that the predictor $\Rin$ now affects the decisions $D$. This emulates the way RAIs are intended to work in practice; for example, a criminal defendant labeled high-risk $(S = 1)$ be a RAI might be less likely to be released pre-trial $(D = 0)$ than they would have been prior to the introduction of the RAI. The data generating process is designed to meet assumptions A1-A3, with $\pi(D \mid A,X,\Rin)$ upper bounded at $0.975$. It is described fully in Appendix \ref{appendix:simulations}. We apply our method to the post-RAI data generating process, simulating the application of post-processing to a predictor that is already embedded in a decision making context. 

\subsubsection{Theorems \ref{thm:risk_gap}-\ref{thm:double_robustness}} \label{subsubsection:simulations1}
To simulate the estimation of the LP coefficient vectors at a particular rate, we add random noise $\epsilon$ of magnitude $o_\Pb(1/n^{1/4})$ to the nuisance parameters $\mu_0$ and $\pi$\footnote{The noise is added on the logit scale to ensure that $\muhat_0, \pihat$ remain in $[0, 1]$, and $\pihat$ is again truncated to 0.975.}. As described above, in general nonparametric settings, regression functions cannot be estimated at $\sqrt{n}$ rates, but they can be estimated at $n^{-1/4}$ rates under relatively weak assumptions \citep{vaart_semiparametric_2002}.

\begin{figure}
    \centering
    \includegraphics[width=\linewidth]{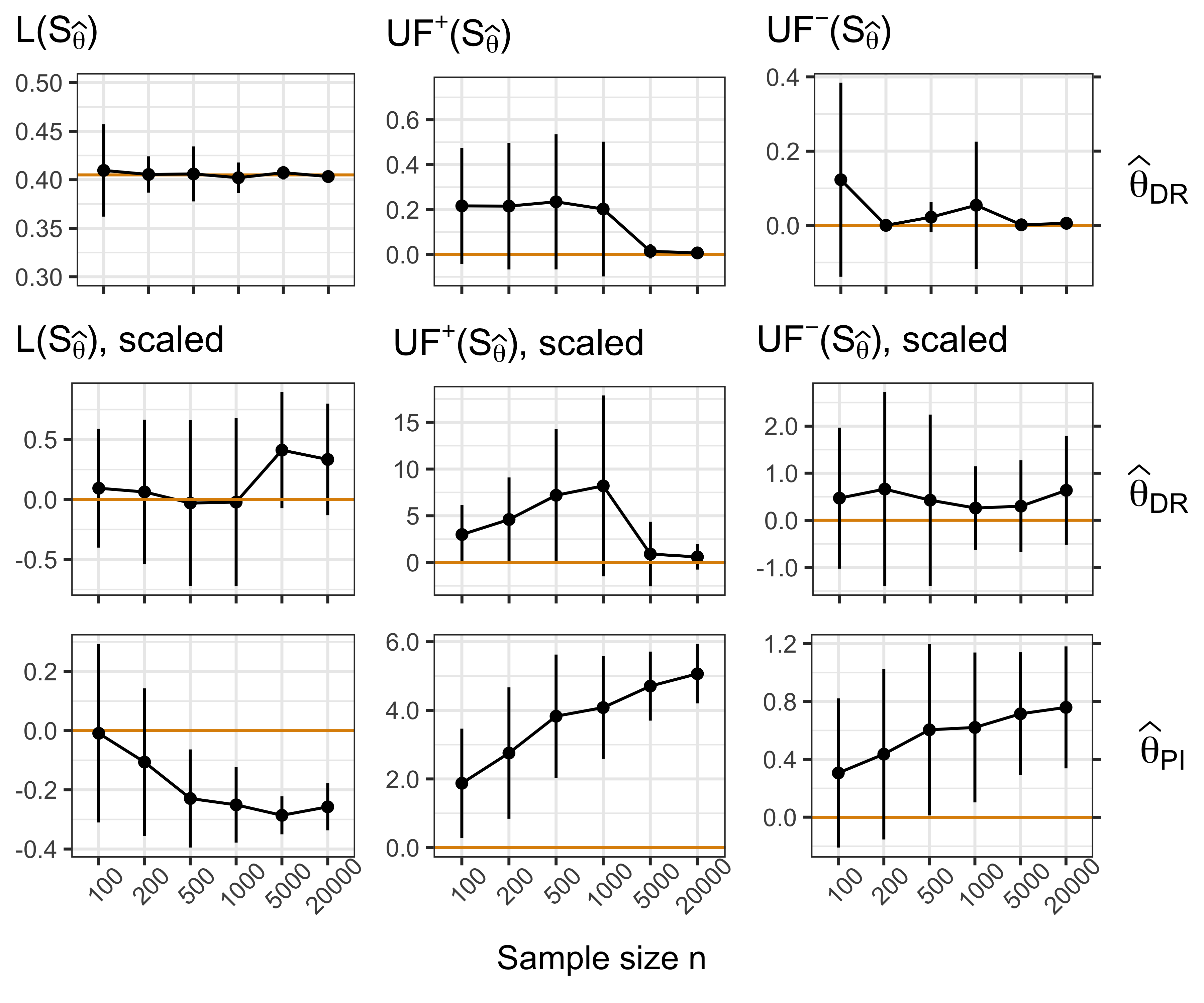}
    \caption{\textbf{(Illustration of Theorems \ref{thm:risk_gap}-\ref{thm:double_robustness})}. Loss $\text{L}(\Rhat)$ and excess unfairness values $\UFP(\Rhat), \UFN(\Rhat)$ for the derived predictor $\Rhat$ for samples of size 100 to 20,000. Each vertical line represents a mean $\pm 1$ sd over 500 simulations. Orange horizontal lines represent the loss of the optimal derived predictor $\Ropt$ (top left panel) or 0. The top row represents our doubly robust (DR) procedure and shows that the loss and excess unfairness converge to their target values. The bottom two rows represent values from the DR procedure or a plugin (PI) procedure, transformed by $\psi(\Rhat) \mapsto \sqrt{n}(\psi(\Rhat) - \psi(\Ropt))$, where $\psi$ is $\risk$ or $\UFP$ or $\UFN$, as appropriate. These rows illustrate that $\sqrt{n}$-convergence is only guaranteed for $\thetahat_{\text{DR}}$: the scaled values for $\thetahat_{\text{DR}}$ do not grow in $n$, while the scaled values for $\thetahat_{\text{PI}}$ begin to diverge.}
    \label{fig:simulations1}
\end{figure}

Figure \ref{fig:simulations1} shows $\risk(\Rhat)$ and the excess unfairness values $\UFP(\Rhat)$, $\UFN(\Rhat)$ for the post-processed predictor $\Rhat$ with fairness constraints $\epsilonpos = 0.10, \-\ \epsilonneg = 0.20$. As expected, when doubly robust estimators are used, the loss and excess unfairness values converge at $\sqrt{n}$ rates to $\risk(\Ropt)$, the loss of the optimal derived predictor and 0, respectively. When plugin estimators are used, the rates are slower than $\sqrt{n}$.

\subsubsection{Fairness-performance tradeoffs} \label{subsubsection:simulations2}
\begin{figure}
    \centering
    \includegraphics[width=\linewidth]{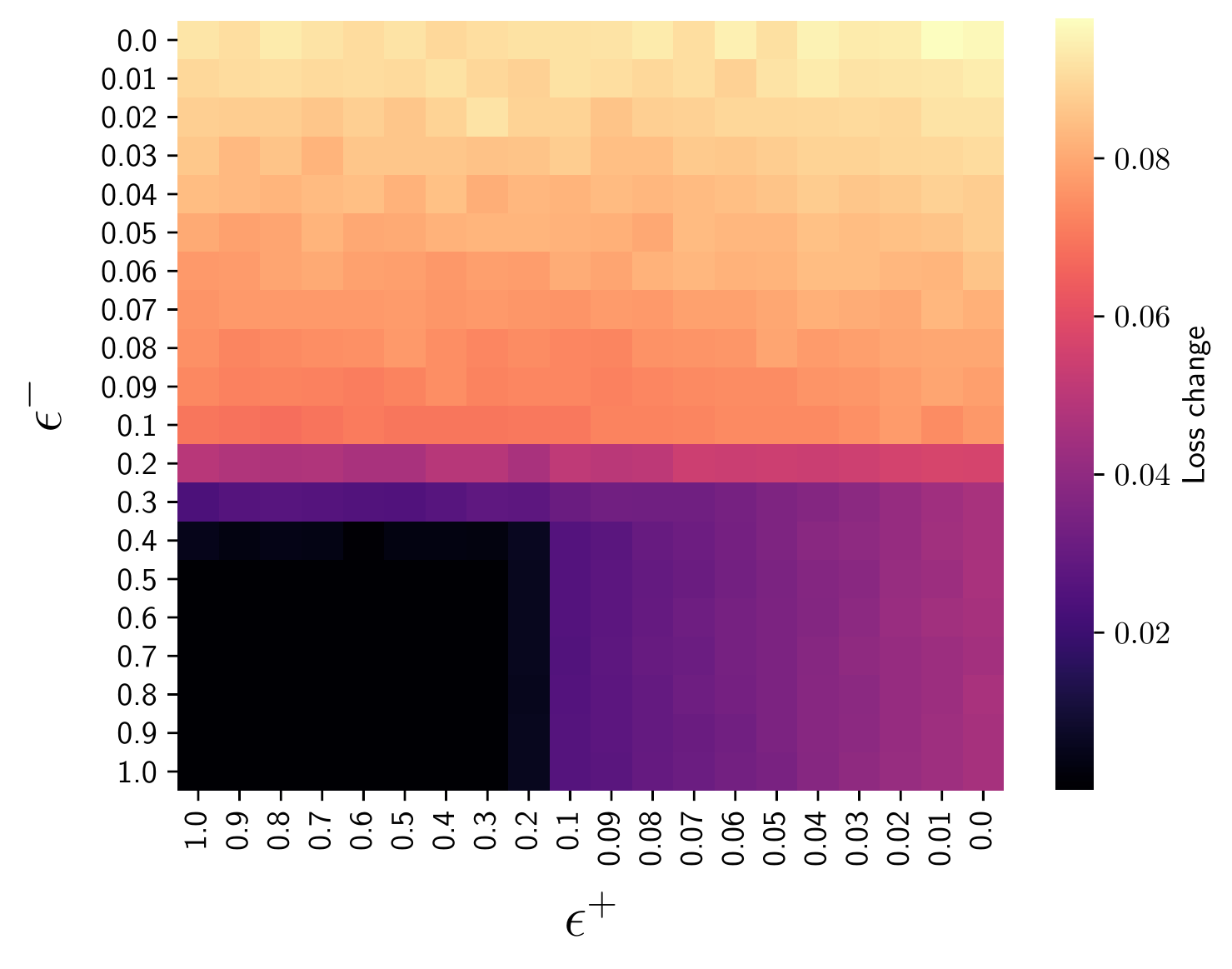}
    \caption{(Fairness-performance tradeoffs). Loss change $\Gamma(\thetaopt) = \risk(\Ropt) - \risk(\Rin)$ for the Bayes-optimal input predictor $\Rin(A, X) = \E[Y^0 \mid A, X]$ and $\thetaopt$ corresponding to different fairness constraints $\epsilonpos, \-\ \epsilonneg$. The black area represents fairness constraints that are looser than the error rate differences of the input predictor $(\Deltapos(\Rin) = 0.24, \-\, \Deltaneg(\Rin) = 0.40)$, which incur no performance cost. The highest performance cost $(0.10)$ occurs when the error rates differences are both constrained to be 0, meaning the derived predictor $\Ropt$ satisfies cEO exactly.}
    \label{fig:risk_fairness_tradeoff}
\end{figure}

Figure \ref{fig:risk_fairness_tradeoff} shows the loss change $\riskchange(\Ropt) = \risk(\Ropt) - \risk(\Rin)$ for each point in a grid of fairness constraints $\epsilonpos, \-\ \epsilonneg$. Here, $\Rin$ is the Bayes-optimal predictor of $Y^0$ in our data generating scenario, meaning $\Rin(A, X) = \E[Y^0 \mid A, X]$. Since any derived predictor necessarily has greater loss than the Bayes-optimal predictor, we refer to the loss change here equivalently as the \emph{performance cost}.

In the data generating process used in the previous section, the Bayes-optimal predictor has absolute error rate differences of only $0.05$ ($\Deltapos$) and $0.04$ ($\Deltaneg$), which leaves little room to illustrate the potential cost of fairness. For these simulations, therefore, we alter the data generating process slightly. (See Appendix \ref{appendix:simulations}). This results in a Bayes-optimal predictor with absolute error rate differences of $0.23$ ($\Deltapos$) and $0.40$ ($\Deltaneg$) and a loss of $0.24$, which are plausible values for a real predictor.

As expected, when $\epsilonpos \geq \Deltapos(\Rin)$ or $\epsilonneg \geq \Deltaneg(\Rin)$, the performance cost is 0: the input predictor already falls satisfies the fairness constraints, so our method simply returns the input predictor. As the tolerances tighten towards 0, the performance declines, though never substantially. For $\epsilonpos = \epsilonneg = 0$, when the derived predictor is constrained to satisfy exact cEO, the loss increases by 0.10, to 0.34. The different values for $\Deltapos(\Rin)$ and $\Deltaneg(\Rin)$ in the input predictor are reflected in the differing costs of satisfying fairness along the two axes: the cost of controlling $\Deltapos(\Rt)$ are lower than the costs of controlling $\Deltaneg(\Rt)$.

\citet{woodworth_learning_2017} showed that post-processing can result in predictors with poor performance, but it is unclear how likely this is to be a problem in practice. While the fairness-accuracy tradeoff naturally depends on the data generating process, our example illustrates that fairness can in some cases be achieved without substantial performance costs.

\subsection{COMPAS data} \label{subsection:compas}
We illustrate our method on the COMPAS recidivism dataset gathered by ProPublica \citep{Angwin2016a, Larson2016a}. COMPAS refers to a collection of tools designed to assess the risk of recidivism. The dataset comprises public arrest records, criminal records, and COMPAS RAI scores from Broward County, Florida, spanning 2013--2016. After filtering the data in the same manner as \citet{Larson2016a} and restricting to defendants who are labeled African-American ($A = 0$) or Caucasian ($A = 1$), we are left with data for 5278 individuals (3175 African-American, 2103 Caucasian).

We utilize the COMPAS scores for general, as opposed to violent, recidivism. The scores are given in risk deciles. Since our method operates on a binary input predictor $\Rin$, we follow ProPublica and set scores of 1-4 to $S = 0$ (``low risk'') and scores of 5-10 to $S = 1$ (``high risk''). The outcome $Y$ is recidivism within a two-year time period. (See \citet{Larson2016a} for how recidivism is operationalized.) ProPublica's analysis focuses on the use of COMPAS to inform pretrial release decisions. The dataset includes dates in and out of jail but does not indicate whether defendants were released pretrial, so we set the treatment $D$ to 0 if defendants left jail within three days of being arrested, and 1 otherwise. This yields 3645 released individuals (2158 African-American, 1487 Caucasian) and 1633 incarcerated individuals (1017 African-American, 616 Caucasian). Note that this threshold is somewhat arbitrary. Florida state law generally requires individuals to be brought before a judge for a bail hearing within 48 hours of arrest, but it may take time for individuals to post bail if they are required and able to do so.

The covariates $X$ consist of gender (coded male or female), age (coded categorically for $< 25$, between 25 and 45, and $> 45$), the number of prior crimes, and charge degree (misdemeanor or felony). Without consulting with domain experts, it is difficult to assess the plausibility of the positivity and ignorability assumptions given these covariates. Hence we intend our analysis primarily to be illustrative of our method, and we resist drawing strong substantive conclusions about COMPAS.  

We weight false positives and false negatives equally, i.e. we set $\wfp = \wfn = 1$. We randomly split the data into training and test sets of equal size. For $\epsilon \in \{0, 0.01, 0.05, 0.10, 0.20, \ldots, 0.90, 1\}$, we set the fairness constraints to $\epsilonpos = \epsilonneg = \epsilon$, compute the corresponding estimate $\thetahat$ on the training set, and estimate properties of the post-processed predictor $\Rhat$ on the test set. We also estimate properties of the binarized COMPAS score $\Rin$ on the test set. We use random forests to estimate both the propensity scores $\pihat$ and the outcome regression $\muhat_0$. To reduce the variance of the estimates, we employ 5-fold cross-fitting: within the train set, we compute five estimates $\thetahat_j, j = 1, \ldots 5$, using four folds at a time to estimate the nuisance parameters and the held-out fold to compute $\thetahat_j$. Then $\thetahat := \frac{1}{5}\sum_j \thetahat_j$. We utilize the test set in an analogous fashion for the remaining estimators.

Table \ref{t:compas} contains estimates and confidence intervals for COMPAS and for the post-processed predictor corresponding to fairness constraints of $\epsilonpos = \epsilonneg = 0.05$. The loss for COMPAS is 0.36, and the differences in the $\cFPR$ and $\cFNR$ are -0.24 and 0.16, respectively. The signs of these differences are consistent with what ProPublica found in their analysis with respect to observable $Y$: the false positive rates are higher for African-American defendants, while the false negative rates are higher for Caucasian defendants. The post-processing procedure successfully shrinks these differences to -0.05 and -0.03, which fall within the target range of $[-0.05, 0.05]$. This reduction corresponds to flipping $9\%$ of the COMPAS scores, and it incurs an increase in risk of only 0.03.

The value of $\thetahat$ corresponding to $\Rhat$ here is $(0, 0.91, 0.23, 1)$. The 0 and the 1 indicate that $\Rhat$ does not change the COMPAS scores for African-American defendants who receive a ``low-risk'' score or Caucasian defendants who receive a ``high-risk'' score. The scores for high-risk African-American defendants are flipped to low-risk $1 - 0.91 = 9\%$ of the time, while the scores for low-risk Caucasian defendants are flipped to high-risk $23\%$ of the time. This has the effect of increasing the false positive rate and decreasing the false negative rate for Caucasians, while moving the rates in the opposite directions for African-Americans.

\begin{figure}
    \centering
    \includegraphics[width=\linewidth]{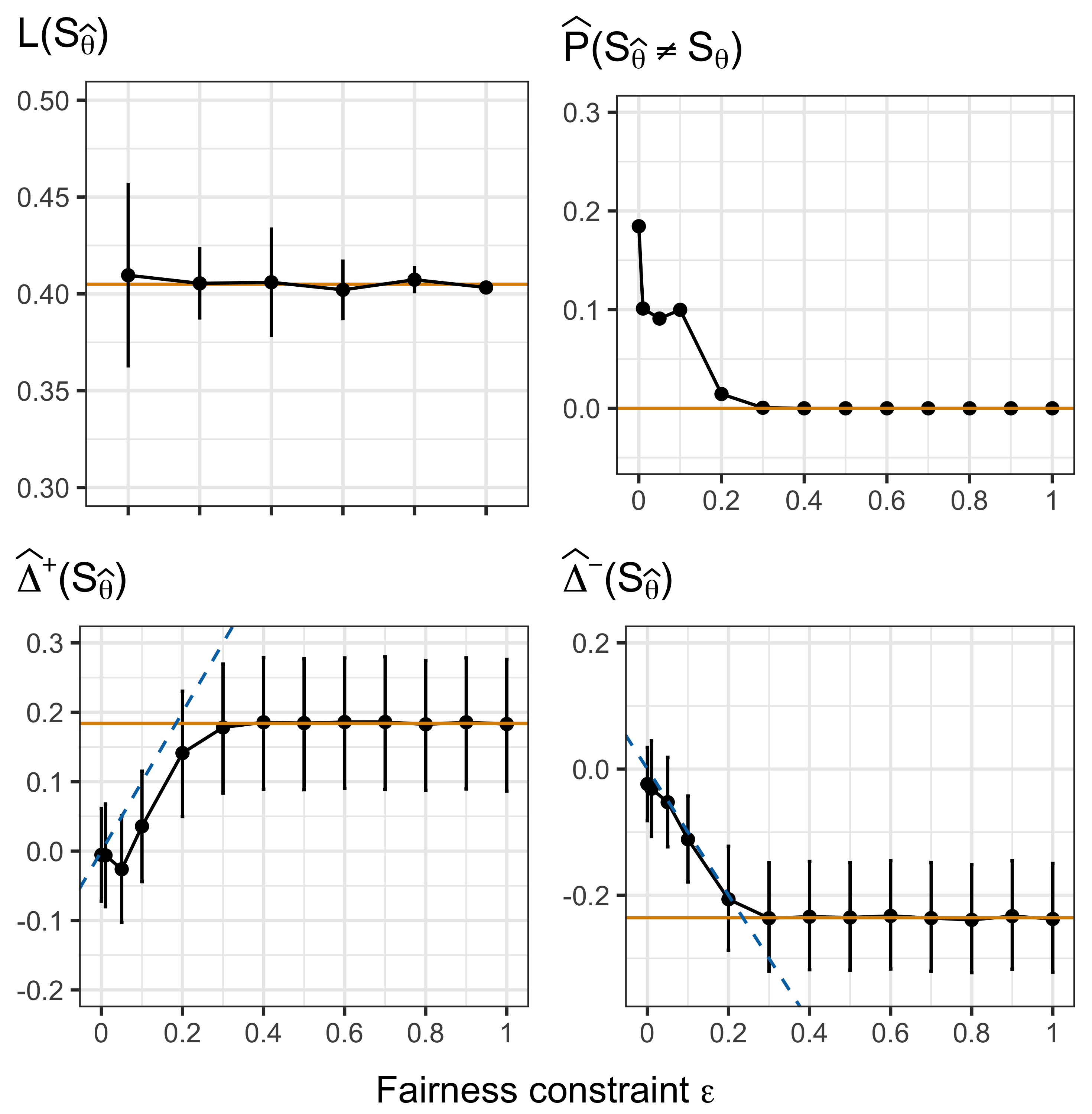}
    \caption{Convergence of the estimated loss $\riskhat(\Rhat)$, predictive change $\widehat{\Pb}(\Rhat \neq \Rin)$, and error rate differences $\Deltahatpos(\Rhat), \Deltahatneg(\Rhat)$, for post-processed versions of the binarized COMPAS predictor. Fairness constraints are set to $\epsilonpos = \epsilonneg = \epsilon$ over a range of values $\epsilon$. Vertical lines are 95\% CIs. Horizontal orange lines indicate the reference values for COMPAS, or 0 in the case of predictive change. The dashed blue lines $y = x$ and $y = -x$, mark the target fairness constraints.}
    \label{f:compas}
\end{figure}

Figure \ref{f:compas} shows the loss, error rate differences, and predictive change for fairness constraints ranging from 0 (requiring no gap in error rates) to 1 (imposing no fairness constraints). Each constraint induces an estimate $\thetahat$ and a corresponding post-processed predictor $\Rhat$. The estimated fairness gaps fall along or within the lines $y = \pm x$, indicating that each $\Rhat$ satisfies its target constraints. At the most stringent setting of 0, the loss for $\Rhat$ is approximately 0.40, which compares favorably with the estimated baseline loss of 0.36 for COMPAS. This $\Rhat$ flips slightly less than $20\%$ of the scores.

For $\epsilon > 0.24$, the fairness constraints are essentially no longer active, since COMPAS itself satisfies these constraints. Indeed, as expected, the $\thetahat$ values for $\epsilon > 0.24$ are all essentially $[0, 1, 0, 1]$, meaning that $\Rhat = \Rin$, and the estimated risk and fairness values all fall close to the estimated values for COMPAS. (There is still some variation in the estimated values due to randomness in the k-fold cross-fitting procedure.)

These results illustrate that our approach performs as intended on a real dataset: if these data were indeed generated from a distribution satisfying the identifying assumptions, then our post-processed predictor would satisfy approximate counterfactual equalized odds while incurring little cost in performance.

{\renewcommand{\arraystretch}{1.2}
\begin{table}
    \centering
    \caption{Estimates and 95\% confidence intervals for the loss $\risk$, loss change $\Gamma$, error rates $\cFPR$ and $\cFNR$ for groups 0 and 1, error rate differences $\Deltapos, \Deltaneg$, and predictive change $\Pb(\Rhat \neq \Rin)$ for the binarized COMPAS predictor $\Rin$ and the post-processed predictor $\Rhat$, with $\epsilonpos$ and $\epsilonneg$ set to $0.05$.}
    \label{t:compas}
        \begin{tabular}{lcc}
        \toprule
        {} &                   $\Rin$ &            $\Rhat$ \\
        \midrule
        $\riskhat(\cdot)$        &     0.36  (0.32, 0.41) &    0.39  (0.35, 0.42) \\
        $\riskchangehat(\cdot)$ &     -- &    0.03  (0.01, 0.04) \\
        \hline
        $\widehat{\cFPR}(\cdot, 0)$        &     0.43  (0.36, 0.49) &    0.39  (0.33, 0.45) \\
        $\widehat{\cFPR}(\cdot, 1)$        &     0.24  (0.18, 0.31) &    0.42  (0.37, 0.47) \\
        $\widehat{\cFNR}(\cdot, 0)$        &     0.30  (0.25, 0.35) &    0.36  (0.31, 0.40) \\
        $\widehat{\cFNR}(\cdot, 1)$        &     0.53  (0.46, 0.60) &    0.41  (0.35, 0.46) \\
        \hline
        $\Deltahatpos(\cdot)$     &  -0.24  (-0.32, -0.15) &  -0.05  (-0.12, 0.02) \\
        $\Deltahatneg(\cdot)$     &     0.18  (0.09, 0.28) &  -0.03  (-0.10, 0.05) \\
        \hline
        $\widehat{\Pb}(\cdot \neq \Rin)$ &     -- &    0.09  (0.09, 0.09) \\
        \bottomrule
        \end{tabular}
\end{table}}

\subsection{Child welfare data}
Cost-sensitive loss functions can drive $\thetahat$ to a trivial classifier that always predicts one class. We illustrate this phenomenon on a dataset representing calls to a child-welfare hotline in Allegheny County, Pennsylvania. The data comprises over 30,000 calls and contains over 1,000 features. The features describe allegations made in the call, assessments of risk made by hotline workers, and features pertaining to individuals associated with the call. Workers must decide whether to \emph{screen in} a call, which means opening an investigation into the allegations. The baseline decision $D = 0$ is to screen out, meaning no investigation takes place. The outcome $Y$ is re-referral to the hotline within a six month period. For further details about the child welfare setting and this dataset in particular, see \citet{Chouldechova2018} and \citet{coston_counterfactual_2020}.

Unlike the COMPAS dataset, this dataset does not include a previously trained predictor. We therefore first build a predictor $\Rin$ that predicts $Y^0$, and then we post-process $S$. In this setting, we have reason to believe that the identification assumptions in section \ref{section:identification} are plausible, once cases with the highest propensity for screen-in are removed; see \citet{coston_counterfactual_2020}. (RAIs are not necessary or useful for cases that are already guaranteed to be screened in.) In order to accomplish this filtering, we first build a propensity score model using random forests on roughly one third of the data. The model appears well-calibrated, so we filter out the approximately 20\% of the cases with estimated propensity scores greater than 0.99. Note that downstream results did not change substantially when these cases were left in.

We then train a classification random forest $\Rin$ to predict $Y$ conditional on $A, X, D = 0$, using the same third of the data. Under the identifying assumptions, $Y|X, A, D = 0$ is equal in distribution to $Y^0|A, X$, so $\Rin$ is indeed an estimate of the target $Y^0$. Following recommended usage in this setting, we set the classification threshold to capture the top 25\% riskiest cases \citep{Chouldechova2018}. 

The predictor $\Rin$ has estimated error rate differences and 95\% confidence intervals of $\Deltahatpos = -0.02 \pm 0.01$ and $\Deltahatneg = 0.09 \pm 0.08$. It is unsurprising that these differences are small, given that rereferral rates are similar for Black (0.24) and White (0.27) cases. See \citet{Chouldechova2017} for an examination of the relationship between base rates and error rates.

\begin{figure}
    \centering
    \includegraphics[width=\linewidth]{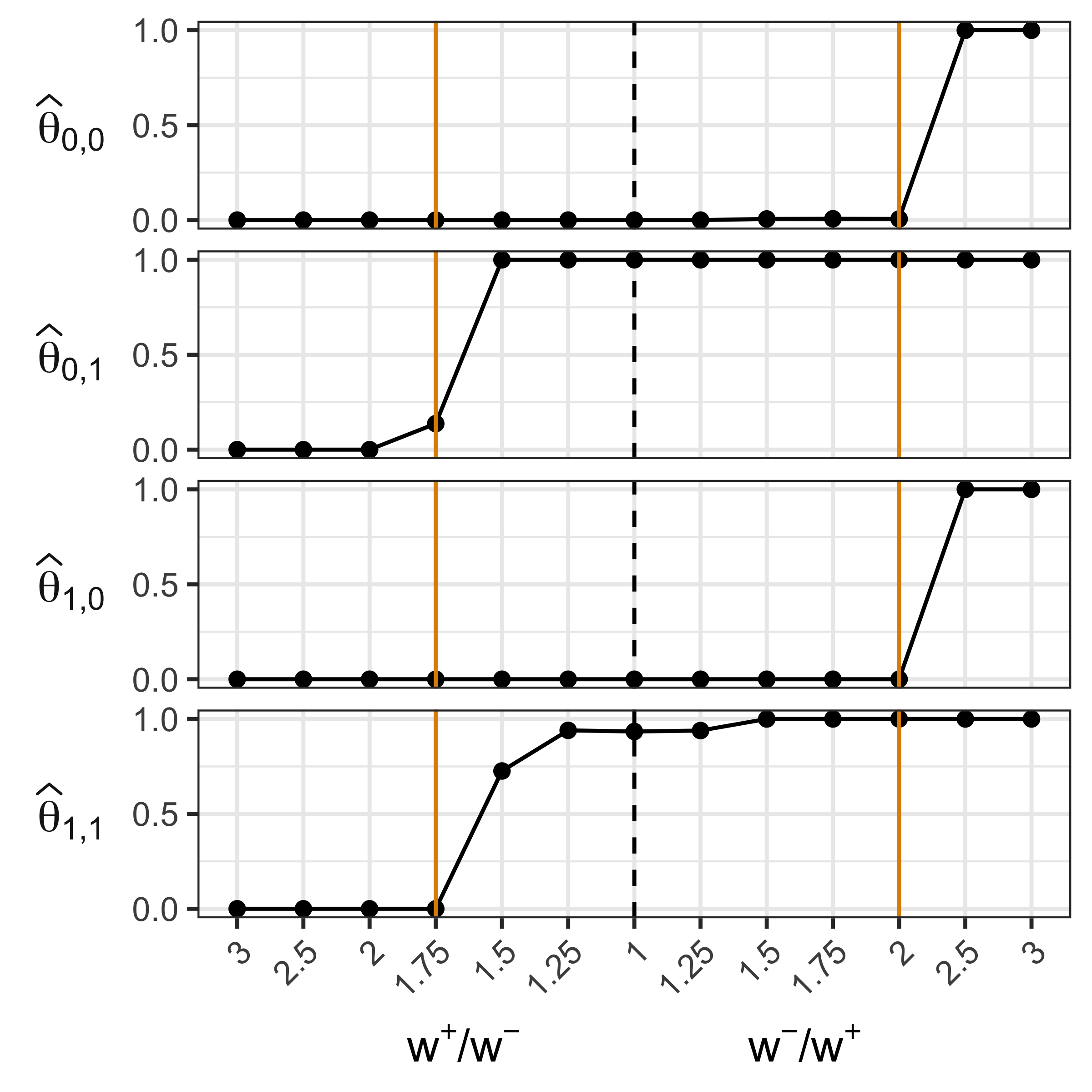}
    \caption{Cost-sensitive post-processing for the child welfare predictor over a range of cost ratios, with fairness constraints $\epsilonpos = \epsilonneg = 0.01$. Each column represents a single $\thetahat$, with the four components $\theta_{a,\rin}$ for $a, \rin \in \{0, 1\},$ in rows. False positives are weighted between 1.25 and 3 times as heavily as false negatives to the left of the dashed line, and vice versa to the right. Extreme cost ratios push the post-processed classifier to a trivial classifier that always predicts 0 (to the left of the orange lines) or 1 (to the right). Between these, post-processing essentially returns the input predictor.}
    \label{f:child_welfare}
\end{figure}

In order to have nontrivial (active) fairness constraints, we set $\epsilonpos = \epsilonneg = 0.01$. Figure \ref{f:child_welfare} shows the value of $\thetahat$ over a range of cost ratios $\wfp/\wfn$ and $\wfn/\wfp$. When false positives are weighted more than 1.5 times as heavily as false negatives, post-processing returns classifiers that are very close to the simple majority classifier $\Rin_{(0, 0, 0, 0)} \equiv 0$. When false negatives are weighted more than 2 times as heavily as false positives, post-processing returns the simple minority classifier $\Rin_{(1, 1, 1, 1)} \equiv 1$. Since the input classifier is approximately fair, between those ranges, post-processing returns classifiers that are very close to the input classifier $\Rin = \Rin_{(0, 1, 0, 1)}$, with only the fourth component $\thetahat_{1,1}$ deviating slightly from 1.

This behavior is expected. Note that a simple majority or minority classifier always satisfies counterfactual equalized odds, since the error rate differences are 0. Since the post-processed predictor only has access at runtime to two binary features, as either false positives or false negatives become sufficiently important, one of these simple classifiers will at some point become the lowest risk option. This is possible in principle when $\wfp$ and $\wfn$ are equal, but it is guaranteed as their ratio grows. 
Since this dataset did not include a pretrained predictor of $Y^0$, it would be preferable to adopt an in-processing approach, i.e. to train a predictor that satisfies the desired fairness constraints in a single stage, rather than training an unconstrained predictor and then post-processing it. We pursue this task in ongoing work.

\section{Discussion and conclusion} \label{section:discussion}
In this paper we considered fairness in risk assessment instruments (RAIs), which are naturally concerned with potential outcomes rather than strictly observable outcomes. We defined the fairness criterion \emph{approximate counterfactual equalized odds} (approximate cEO), which allows users to negotiate the tradeoff between fairness and performance. We argued that this fairness criterion is likelier than other candidate criteria to reduce discriminatory disparate impact, which we defined as $D \not\ind A \mid Y^0$.

We presented a method to post-process an existing binary predictor to satisfy approximate cEO using doubly robust estimators, and we showed that our method has favorable convergence properties. Our rate results translate readily to the post-processing setting of \citep{hardt_equality_2016}, in which the outcome of interest is the observable $Y$ and the fairness criterion is (approximate) observational equalized odds.

Once it is constructed, the post-processed predictor requires access at runtime only to the sensitive feature and the input predictor, making it relatively feasible to implement on top of existing RAIs. A predictor trained from scratch would be constrained by the set of covariates available in deployment, whereas the post-processing approach allows researchers to devise a set of suitable deconfounding covariates and then collect an appropriate dataset on a one-time basis.

In closing, we note that from our perspective, notions of fairness in predictive systems ought to be subordinate to notions of fairness grounded in the actual decisions or events that those systems inform, and the impact that those decisions have on people's lives. Though little is currently known about how decision makers respond to RAIs, there is some evidence that judges do not have much faith in recidivism predictions and that RAIs can have little impact on decisions \citep{jonnson_influence_2018, stevenson_assessing_2018}. As RAIs and the general public's understanding of how they function co-evolve, it is likely that the ways in which decision makers respond to them will evolve as well.

Nevertheless, it seems plausible that some fairness criteria for RAIs are likelier than others to lead to increased (un)fairness with respect to decisions and outcomes. While this is ultimately an empirical question, we believe that this kind of consideration ought to ground discussions of fairness in RAIs and predictive systems generally. As long as there are domains involving high stakes decisions that we do not wish to fully automate, RAIs will remain relevant, and so will the task of ensuring that they lead to a society that is more fair, not less. 


\begin{acks}
Edward Kennedy gratefully acknowledges support from NSF DMS Grant 1810979. Additionally, we are grateful to the Block Center for Technology and Society at Carnegie Mellon University for funding this work, and to the Allegheny County Department of Human Services for furnishing the child welfare data.
\end{acks}
\bibliographystyle{ACM-Reference-Format}
\bibliography{references.bib}

\appendix

\section{Proofs of propositions} \label{appendix:propositions}
For convenience, we restate all our assumptions.
\begin{align*}
    & \text{A1. (Consistency) } \quad Y = DY^1 + (1-D)Y^0 \\
    & \text{A2. (Positivity) } \qquad \exists \delta \in (0, 1): \Pb(\pi(A, X, \Rin) \leq 1 - \delta) = 1 \\
    & \text{A3. (Ignorability) } \, \quad Y^0 \ind D \mid A, X, S \\
    & \text{A4. (Bounded propensity estimator)} \\
    & \quad \quad \exists \gamma \in (0, 1) \text{ s.t. } \Pb(\pihat(A, X, \Rin) \leq 1 - \gamma) = 1 \\
    & \text{A5. (Nuisance estimator consistency)} \\
    & \quad \quad \| \muhat_0 - \mu_0 \| = o_\Pb(1), \-\ \| \pihat - \pi \| = o_\Pb(1) \\
    & \text{A6. (Nuisance estimator rates).} \\
    & \quad \quad  \Vert \muhat_0 - \mu_0 \Vert \Vert \pihat - \pi \Vert = o_\Pb(1/\sqrt{n})
\end{align*}

\subsection*{Proof of Proposition 1 (Identification of error rates for input predictor $\Rin$)}
\begin{proof}

\begin{align*}
    \cFPR(\Rin, a) &= \Pb(\Rin = 1 \mid Y^0 = 0, A = a) \\
                  &= \frac{\Pb(\Rin = 1, Y^0 = 0, A = a)}{\Pb(Y^0 = 0, A = a)} \\
                  &= \frac{\E[\Rin(1 - Y^0)\one\{A=a\}]}{\E[(1-Y^0)\one\{A=a\}]} \\
                &= \frac{\E[\Rin(1 - \E[Y^0 \mid A, X, \Rin, D=0)]\one\{A=a\}}{\E[(1-\E[Y^0 \mid A,X, \Rin, D = 0])\one\{A=a\}]} \\
                  &= \frac{\E[\Rin(1 - \mu_0)\one\{A=a\}]}{\E[(1-\mu_0)\one\{A=a\}]}
\end{align*}
\begin{align*}
    \cFNR(\Rin, a) &= \Pb(\Rin = 0 \mid Y^0 = 1, A = a) \\
                  &= \frac{\Pb(\Rin = 0, Y^0 = 1, A = a)}{\Pb(Y^0 = 1, A = a)} \\
                  &= \frac{\E[(1-\Rin)Y^0\one\{A=a\}]}{\E[Y^0\one\{A=a\}]} \\
                &= \frac{\E[(1-\Rin)\E[Y^0 \mid A, X, \Rin, D=0]\one\{A=a\}}{\E[\E[Y^0 \mid A,X, \Rin, D = 0]\one\{A=a\}]} \\
                  &= \frac{\E[(1-\Rin)\mu_0\one\{A=a\}]}{\E[\mu_0\one\{A=a\}]}
\end{align*}
The fourth equality in both derivations uses iterated expectation as well as positivity and ignorability, and the fifth equality uses consistency.
\end{proof}


\subsection*{Proof of Proposition 2 (Identification of the loss and fairness constraints)}
\begin{proof}
Considering just the first component of the loss, we have:
\begin{align*}
    (\wfp)\Pb(\Rt = 1, Y^0 = 0) &= (\wfp)\E[\Rt(1 - Y^0)] \\
    &= (\wfp)\E[\E[\Rt(1-Y^0)|A,S,X]] \\
    &= (\wfp)\E\left\{\E[\Rt|A,\Rin](1 - \E[Y^0|A,X,\Rin])\right\} \\
    &= (\wfp)\E\left\{\theta_{A,\Rin}(1 - \E[Y^0|A,X,\Rin, D = 0])\right\} \\
    &= (\wfp)\E\left\{\theta_{A,\Rin}(1 - \mu_0)\right\}
\end{align*}
where the third equality uses that $\Rt$ only depends on $(A, S)$, the fourth uses the definition of $\theta_{A,\Rin}$ and ignorability, and the fifth uses consistency. Similar reasoning shows that $(\wfn)\Pb(S = 0, Y^0 = 1) = (\wfn)\E\left\{(1 - \theta_{A,\Rin})\mu_0\right\}$. Combining these, we have
\begin{align*}
    \risk(\Rt) :&= \wfp\Pb(\Rin = 1, Y^0 = 0) + \wfn\Pb(\Rin = 0, Y^0 = 1) \\
    &= \E[\theta_{A,\Rin}(\wfp - (\wfp + \wfn)\mu_0)] + (\wfn)\E[\mu_0] \\
    &= \theta^T\beta + (\wfn)\E[\mu_0]
\end{align*}
We turn now to the fairness constraints. The error rates of the derived predictor $\Rt$ depend on the error rates on the input predictor $\Rin$ as follows. Beginning with $\cFPR(\Rt, a)$, we have:
\begin{align*}
    & \Pb(\Rt = 1 \mid Y^0 = 0, A=a) = \\
    & \quad \sum_{\rin\in\{0,1\}} \Pb(\Rt = 1 \mid Y^0 = 0, A=a, \Rin = \rin)\Pb(\Rin = \rin \mid Y^0 = 0, A=a) \\
    \-\ =& \quad \sum_{\rin\in\{0,1\}} \Pb(\Rt = 1 \mid A=a, \Rin=\rin)\Pb(\Rin = \rin \mid Y^0 = 0, A=a) \\
    \-\ =& \quad \theta_{a, 0} (1 - \cFPR(\Rin, a)) + \theta_{a, 1}\cFPR(\Rin, a)
\end{align*}
where the first equality simply involves conditioning on $\Rin$, and the second equality uses that $\Rt \ind Y^0 \mid A, \Rin$. In other words, the false positive rate of $\Rt$ depends only on $\theta$ and the false positive rate of the input predictor $\Rin$. For the cFNR, by similar reasoning, we have:
\begin{align*}
    &\Pb(\Rt = 0 \mid Y^0 = 1, A=a) = \\
    & \quad 1 -\theta_{a,0}(\cFNR(\Rin, a)) + \theta_{a,1}(\cFNR(\Rin, a) - 1)
\end{align*}
The identification statements in the proposition follow by simply substituting in the expressions for $\cFPR(\Rin, a), \cFNR(\Rin, a)$ from Proposition 1 and rearranging.
\end{proof}

For any function $f(A, X, \Rin)$, it straightforward to show that
\begin{align*}
    \E[f(A, X, \Rin)\phi] &= \E[f(A, X, \Rin)\mu_0]
\end{align*}
It follows that the identification results in Propositions \ref{proposition:identification_input} and \ref{proposition:identification_derived} hold when $\mu_0$ is replaced by $\phi$. We utilize this equivalence in the proofs of Theorems \ref{thm:double_robustness} and \ref{thm:asymptotic_normality}.

\section{Proofs of Theorems} \label{appendix:theorems}


\subsection{Theorem 1 (Loss gap)}

We first introduce a lemma used in the proof of the theorem. The lemma gives sufficient conditions under which the optimal value of an estimated convex program converges at a particular rate $f(n)$ to the optimal value of the target convex program. It is a adaptation of Theorem 3.5 in Shapiro (1991) that follows immediately from Theorems 2.1 and 3.4 in that same paper.

\begin{lemma}[Shapiro, 1991] \label{lemma:shapiro}
Let $\Theta$ be a compact subset of $\R^k$. Let $C(\Theta)$ denote the set of continuous real-valued functions on $\Theta$, with $\mathscr{L} = C(\Theta) \times \ldots \times C(\Theta)$ the $r$-dimensional Cartesian product. Let $\psi(\theta) = \left(\psi_{0}, \ldots, \psi_{r}\right) \in \mathscr{L}$ be a vector of convex functions. Consider the quantity $\alpha^*$ defined as the solution to the following convex optimization program: 
\begin{align*}
    \alpha^* = \min_{\theta\in\Theta} \quad & \psi_0(\theta) \\
              \text{subject to }
                 & \psi_j(\theta) \leq 0, \-\ j = 1, \ldots, r
\end{align*}
Assume that Slater's condition holds, so that there is some $\theta \in \Theta$ for which the inequalities are satisfied and non-affine inequalities are strictly satisfied, i.e. $\psi_j(\theta) < 0$ if $\psi_j$ is non-affine. Now consider a sequence of approximating programs, for $n = 1, 2, \ldots$:
\begin{align*}
    \widehat{\alpha}_n = \min_{\theta\in\Theta} \quad & \psihat_{0n}(\theta) \\
              \text{subject to }
                 & \psihat_{jn} (\theta) \leq 0, \-\ j = 1, \ldots, r
\end{align*}
with $\psihat_n(\theta) := \left(\psihat_{0n}, \ldots, \psihat_{rn}\right) \in \mathscr{L}$. Assume that $f(n)(\psihat_n - \psi)$ converges in distribution to a random element $W \in \mathscr{L}$ for some real-valued function $f(n)$. Then:
\begin{align*}
    f(n)(\widehat{\alpha}_n - \alpha_0) \rightsquigarrow L
\end{align*}
for a particular random variable $L$. It follows that $\widehat{\alpha}_n - \alpha_0 = O_\Pb(1/f(n))$.
\end{lemma}

\subsubsection{Proof of theorem}
We expand the loss by introducing the term $\betahat^T\thetahat$, which is the quantity that is minimized in the course of computing $\thetahat$. We proceed by splitting the loss into two terms and showing that each of those terms is $O_\Pb(1/f(n))$.

\begin{proof}
The loss gap can be expanded as follows:
\begin{align*}
    \risk(\Rhat) - \risk(\Ropt) &= \beta^T\thetahat - \beta^T\thetaopt \\
    &= \underbrace{\left(\beta^T\thetahat - \betahat^T\thetahat\right)}_{(1)} + \underbrace{\left(\betahat^T\thetahat - \beta^T\thetaopt\right)}_{(2)}
\end{align*}
For term (1), we have
\begin{align*}
    \thetahat^T\left(\beta - \betahat\right) &\leq \Vert \thetahat \Vert \Vert \beta - \betahat \Vert  \\
    &\leq 2\Vert \beta - \betahat \Vert \\
    &= O_\Pb(1/f(n))
\end{align*}
where the first line uses Cauchy-Schwarz, the second line follows from the fact that $\thetahat \in [0, 1]^4$, and the third line follows by assumption. For term (2), we rely on Lemma \ref{lemma:shapiro}. Note that we can write
\begin{align*}
    \risk(\Ropt) = \min_{\theta\in\Theta} \quad & \psi_0(\theta) \\
              \text{subject to }
                 & \psi_j(\theta) \leq 0, \-\ j = 1, \ldots, 4 \\
    \riskhat(\Rhat) = \min_{\theta\in\Theta} \quad & \psihat_0(\theta) \\
              \text{subject to }
                 & \psihat_j(\theta) \leq 0, \-\ j = 1, \ldots, 4
\end{align*}
with $\Theta = [0, 1]^4$, and $\psi(\theta) = (\psi_0(\theta), \ldots, \psi_4(\theta))$ defined by
\begin{align*}
    \psi(\theta) &= (\risk, \-\ \Deltapos - \epsilonpos, \-\ -\Deltapos - \epsilonpos, \-\ \Deltaneg - \epsilonneg, \-\ -\Deltaneg - \epsilonneg) \\
    \psihat(\theta) &= (\riskhat, \-\ \Deltahatpos - \epsilonpos, \-\ -\Deltahatpos - \epsilonpos, \-\ \Deltahatneg - \epsilonneg, \-\ -\Deltahatneg - \epsilonneg)
\end{align*}
where for brevity we omit the argument $\Rt$ to $\risk$ and the error rate differences $\Deltapos, \Deltaneg$. Since these are linear programs, Slater's condition is satisfied. (The LPs are always feasible, since $(0, 0, 0, 0)$ and $(1, 1, 1, 1)$ are always solutions.) By assumption, each of the estimators in $\psihat(\theta)$ converges at rate $f(n)$, so $f(n)\left(\psihat(\theta) - \psi(\theta)\right)$ converges to some (unknown) random variable. (We rule out pathological cases in which this does not happen.) Per Lemma \ref{lemma:shapiro}, it follows that $\betahat^T\thetahat - \beta^T\thetaopt = \riskhat(\Rhat) - \risk(\Ropt) = O_\Pb(1/f(n))$. 

The sum of the two terms in the loss gap is therefore also $O_\Pb(1/f(n))$.
\end{proof}

\subsection{Theorem 2 (Excess unfairness)}
The proof relies on the following lemma, as well as the convergence of the estimated LP coefficient vectors $\betaposhat, \betaneghat$. When $\betaposhat, \betaneghat$ are close to $\betapos, \betaneg$, the excess unfairness must be small for any $\theta \in \Theta = [0, 1]^4$, including of course $\thetahat$. 

\begin{lemma} \label{lemma:rate_bounds}
Let $\xi, W$ be constant vectors and $\widehat{\xi}_n, \widehat{W}_n$ be random vectors, with $\Vert \xi - \widehat{\xi}_n \Vert = O_\Pb(1/f(n))$ for some real-valued $f(n)$. If, for all $M > 0$, $\Pb(\Vert W - \widehat{W}_n \Vert > M) \leq \Pb(\Vert \xi - \widehat{\xi}_n \Vert > CM)$ for some constant $C$, then $\Vert W - \widehat{W}_n \Vert = O_\Pb(1/f(n))$.
\begin{proof}
For any $\epsilon > 0$, there exists some $M_\epsilon > 0$ such that $\Pb(f(n)\Vert \xi - \widehat{\xi}_n \Vert > M_\epsilon) < \epsilon$  for all $n$ large enough. Set $M = M_\epsilon/C$. Then $\Pb(f(n)\Vert W - \widehat{W}_n \Vert > M) < \epsilon$ for all $n$ large enough.
\end{proof}
\end{lemma}

\subsubsection{Proof of theorem}
\begin{proof}
Fix $\delta \in (0, 1)$. We have
\begin{align*}
    & \Pb\left(\UFP(\Rhat) > \delta \text{ or } \UFN(\Rhat) > \delta\right) \\
    & \leq \Pb\Big( | \theta^T\betapos |  -  | \theta^T\betaposhat |  > \delta \text{ or }  | \theta^T\betaneg |  -  | \theta^T\betaneghat |  > \delta \\
    & \hspace{17em} \text{ for some } \theta \in [0,1]^4\Big) \\
    & \leq \Pb\Big( | \theta^T\betapos - \theta^T\betaposhat |  > \delta \text{ or }  | \theta^T\betaneg - \theta^T\betaneghat |  > \delta \\
    & \hspace{17em} \text{ for some } \theta \in [0,1]^4\Big) \\
    & \leq \Pb\Big(\Vert\theta\Vert\cdot\Vert\widehat{\betapos} - \betapos\Vert > \delta \text{ or } \Vert\theta\Vert\cdot\Vert\widehat{\betaneg} - \betaneg\Vert > \delta \\
    & \hspace{17em} \text{ for some } \theta \in [0,1]^4\Big)
\end{align*}
\begin{align*}
    & \leq \Pb\left(2\Vert\widehat{\beta}^+ - \betapos\Vert > \delta \text{ or } 2\Vert\widehat{\beta}^- - \betaneg\Vert > \delta\right) \\
    & \leq \Pb\left(2\Vert\widehat{\beta}^+ - \betapos\Vert  > \delta\right) + \Pb\left(2\Vert\widehat{\beta}^- - \betaneg\Vert > \delta\right) \\
    & = \Pb\left(\Vert\widehat{\beta}^+ - \betapos\Vert  > \delta/2\right) + \Pb\left(\Vert\widehat{\beta}^- - \betaneg\Vert > \delta/2\right)
\end{align*}
where the third inequality uses Cauchy-Schwartz, the fourth uses that $\theta \in [0, 1]^4 \implies \Vert\theta\Vert \leq 2$, and the fifth uses the union bound. The reasoning in the first inequality is as follows: if $\UFP(\Rhat) > \delta$, then $|\thetahat^T\betapos| - |\thetahat^T\betaposhat| > \delta$, since $\thetahat^T\betaposhat \leq \epsilonpos$ by construction. A necessary condition, then, is that $|\theta^T\betapos| - |\theta^T\betaposhat| > \delta$ for some $\theta \in [0, 1]^4$.

Since $\betaposhat$ and $\betaneghat$ are consistent at rate $f(n)$ by assumption, it follows from Lemma \ref{lemma:rate_bounds} that
\begin{align*}
    \max\left\{\UFP(\Rhat), \UFN(\Rhat)\right\} = O_\Pb(1/f(n))
\end{align*}
\end{proof}

\subsection{Theorem \ref{thm:double_robustness} (Double robustness.)}

Recall that $\Pn(f(Z)) = n^{-1}\sum_{i=1}^n f(Z_i)$ denotes the sample average of any fixed function $f: \mathcal{Z} \mapsto \R$. In this proof and the proof of Theorem \ref{thm:asymptotic_normality}, we let $\Pb(f) = \int f(z)d\Pb(z)$ denote the expected value of a fixed function $f(Z)$ with respect to $Z$. For example, $\Pb(\phihat) = \int \phihat(z) d\Pb(z)$ is the expected value of $\phihat(Z)$ conditional on the sample used to construct $\phihat$.

The proofs of each of these theorems utilize the following two lemmas.

\begin{lemma} \label{lemma:double_robustness}
Let $W$ be a function of (at most) $A, X, S$ such that $\Vert W \Vert \leq M < \infty$ for some $M$. Suppose that $\Vert \muhat_0 - \mu_0 \Vert \Vert \pihat - \pi \Vert = O_\Pb(g(n))$ for some function $g(n)$. Then, under assumption A4 (bounded propensity estimator), 
\begin{align*}
    \Pb\left(W(\phihat - \phi)\right) = O_\Pb(g(n))
\end{align*}
\begin{proof}
\begin{align*}
    \Pb\left(W(\phihat - \phi)\right) &=\Pb\left(W\left(\frac{1-D}{1-\pihat}(Y-\muhat_0) + \muhat_0 - \frac{1-D}{1-\pi}(Y-\mu_0) - \mu_0\right)\right) \\
     &= \Pb\left(W\left(\frac{1-D}{1-\pihat}(\mu_0-\muhat_0) + \muhat_0 - \frac{1-D}{1-\pi}(\mu_0-\mu_0) - \mu_0\right)\right) \\
     &= \Pb\left(W\left(\frac{1-\pi}{1-\pihat}(\mu_0-\muhat_0) + \muhat_0 - \mu_0\right)\right) \\
     &= \Pb\left(W\left(\frac{(\mu_0-\muhat_0)(\pihat-\pi)}{1-\pihat}\right)\right) \\
     &\leq \frac{1}{\gamma}\Pb(W(\mu_0-\muhat_0)(\pihat-\pi)) \\
     &\leq \frac{1}{\gamma}\Vert W\Vert \Vert \mu_0 - \muhat_0 \Vert \Vert \pihat - \pi \Vert \\
     &= O_\Pb(g(n))
\end{align*}
where $\gamma$ is the bound on the propensity estimator in assumption A4. The second and third lines use iterated expectation and consistency; the third line uses iterated expectation, conditioning on $(A, \Rin)$; the fifth line uses assumption A4; and the sixth line uses the Cauchy-Schwarz inequality.
\end{proof}
\end{lemma}

The next lemma is a restatement of Lemma 2 in \citet{kennedy_sharp_2020}.

\begin{lemma}[Kennedy, 2020] \label{lemma:kennedy}
\begin{align*}
    (\Pn - \Pb)(\phihat - \phi) &= O_\Pb\left(\frac{\Vert \phihat - \phi \Vert}{\sqrt{n}}\right)
\end{align*}
\end{lemma}

\subsubsection{Proof of the theorem}
Recall that in the statement of the theorem, $g(n)$ is the convergence rate of $\Vert \mu_0 - \muhat_0 \Vert \Vert \pihat - \pi \Vert$, i.e. $\Vert \mu_0 - \muhat_0 \Vert \Vert \pihat - \pi \Vert = O_\Pb(g(n))$.
\begin{proof}
Note that for a fixed length vector $v \in \R^k$:
\begin{align*}
    \Vert \widehat{v} - v \Vert = O_\Pb(f(n)) \iff \widehat{v}_j - v_j = O_\Pb(f(n)), \-\ j = 1, \ldots k
\end{align*}
It therefore suffices to show that the rate result in the theorem holds for each component of $\betahat, \betaposhat, \betaneghat$.

Starting with $\betahat_{a,\rin}$, a component of $\betahat$, we have the following, by simple addition and subtraction of measures:
\begin{align*}
    \betahat_{a,\rin} - \beta_{a, \rin} = & \-\ (\Pn - \Pb)\left\{\one\{A = a, \Rin = \rin\}(\wfp - (\wfp + \wfn)\phi)\right\} \-\ + \\
    & \-\ (\Pn - \Pb)\left\{\one\{A = a, \Rin = \rin\}(\wfn - \wfp)(\phihat - \phi)\right\} \-\ + \\
    & \-\ \Pb\left\{\one\{A = a, \Rin = \rin\}(\wfn - \wfp)(\phihat - \phi)\right\}
\end{align*}
The first term is $O_\Pb(1/\sqrt{n})$ by the central limit theorem. The second term is $O_\Pb(g(n)/\sqrt{n})$ by Lemmas \ref{lemma:double_robustness} and \ref{lemma:kennedy}. The third term is $O_\Pb(g(n))$ by Lemma \ref{lemma:double_robustness}. Thus
\begin{align*}
    \betahat_{a, \rin} - \beta_{a, \rin} &= O_\Pb\left(\max\{n^{-1/2}, g(n)\}\right)
\end{align*}
and the result therefore holds for $\Vert \betahat - \beta \Vert$.

We now turn to $\betaposhat$ and $\betaneghat$. It suffices to show that the rate result holds for $\widehat{\cFPR}(\Rin, a)$ and $\widehat{\cFNR}(\Rin, a)$, for $a \in \{0, 1\}$. For notational convenience, let $\gamma_a = (1 - \phi)\one\{A = a\}$ and $\gammahat_a = (1 - \phihat)\one\{A = a\}$. We have
\begin{spreadlines}{1em}
\begin{align}
    &\widehat{\cFPR}(\Rin, a) - \cFPR(\Rin, a) = \frac{\Pn[\Rin\gammahat_a]}{\Pn[\gammahat_a]} - \frac{\Pb[\Rin\gamma_a]}{\Pb[\gamma_a]} \nonumber \\
    &= \frac{\Pn[\Rin\gammahat_a]\Pb[\gamma_a] - \Pb[\Rin\gamma_a]\Pn[\gammahat_a]}{\Pn[\gammahat_a]\Pb[\gamma_a]} \nonumber \\
    &= \frac{\Pb[\gamma_a]\Big(\Pn[\Rin\gammahat_a] - \Pb[\Rin\gamma_a]\Big) - \Pb[\Rin\gamma_a]\Big(\Pn[\gammahat_a] - \Pb[\gamma_a]\Big)}{\Pn[\gammahat_a]\Pb[\gamma_a]} \nonumber \\
    &= \Pn[\gammahat_a]^{-1}\Big\{\underbrace{\left(\Pn[\Rin\gammahat_a] - \Pb[\Rin\gamma_a]\right)}_{(1)} - \cFPR(\Rin, a)\underbrace{(\Pn[\gammahat_a] - \Pb[\gamma_a])}_{(2)}\Big\} \label{eq:unfairness_expansion}
\end{align}
\end{spreadlines}
The two terms can be expanded as follows:
\begin{align*}
    (1) &= (\Pn - \Pb)\Rin\gamma_a + (\Pn - \Pb)(\Rin(\gammahat_a - \gamma_a)) + \Pb(\Rin(\gammahat_a - \gamma_a)) \\
    (2) &= (\Pn - \Pb)\gamma_a + (\Pn - \Pb)(\gammahat_a - \gamma_a) + \Pb(\gammahat_a - \gamma_a)
\end{align*}
Once again, in both these expressions the first term is $O_\Pb(1/\sqrt{n})$ by the central limit theorem, the second term is $O_\Pb(g(n)/\sqrt{n})$ by Lemma \ref{lemma:kennedy}, and the third term is $O_\Pb(g(n))$. Under assumption A4 (bounded propensity estimator), $\Pn[\gammahat_a]^{-1}$ is bounded a.s., and $\cFPR(\Rin, a)$ is always bounded in $[0, 1]$. Therefore, we can rewrite \eqref{eq:unfairness_expansion} as
\begin{align}
    & \Pn[\gammahat_a]^{-1}(\Pn - \Pb)\Big\{\big(\Rin - \cFPR(\Rin, a)\big)\gamma_a\Big\} + O_\Pb(g(n)) \label{eq:unfairness_expansion_final}
\end{align}
This expression is $O_\Pb(\max\{n^{-1/2}, g(n)\}$ and therefore so is $\Vert \betaposhat - \betapos \Vert$. The result for $\widehat{\cFNR}(\Rin, a)$, and consequently for $\Vert \betaneghat - \betaneg \Vert$, follows by identical reasoning, with $\gamma_a$ redefined as $\phi\one\{A = a\}$ so that $\cFNR(\Rin, a) = \E[(1 - S)\gamma_a]/\E[\gamma_a]$.
\end{proof}

\subsection{Theorem \ref{thm:asymptotic_normality} (asymptotic normality)}
For ease of reference, we reiterate the following quantities defined in the theorem.
\begin{align*}
    f_\theta &= \theta_{A,\Rin}(\wfp - (\wfp + \wfn)\phi) + \wfn\phi \\
    g_a &= (\theta_{a,1} - \theta_{a,0})\frac{\one\{A=a\}(1 - \phi)(\Rin - \cFPR(\Rin, a))}{\E[\one\{A=a\}(1 - \phi)]} \\
    h_a &= (\theta_{a,1} - \theta_{a,0})\frac{\one\{A=a\}\phi(\Rin - \cFNR(\Rin, 0))}{\E[\one\{A=a\}\phi]}
\end{align*}
Define $\fhat_\theta, \-\ \ghat_a, \-\ \hhat_a$ analogously, substituting $\phihat, \-\ \widehat{\cFPR}, \-\ \widehat{\cFNR}$ for $\phi$, $\cFPR, \-\ \cFNR$.

We first prove the statements for the loss $\risk$ and loss change $\riskchange$. Note that $\risk(\Rt) = \Pb[f_\theta]$ and $\riskhat(\Rt) = \Pn[\fhat_\theta]$. By simple addition and subtraction of measures, we have
\begin{align*}
    \riskhat(\Rt) - \risk(\Rt) &= (\Pn - \Pb)f_\theta + (\Pn - \Pb)(\fhat_\theta - f_\theta) + \Pb(\fhat_\theta - f_\theta)
\end{align*}
Under assumption A6 (nuisance estimator rates), the third term in this sum is $o_\Pb(1/\sqrt{n})$ by Lemma \ref{lemma:double_robustness}, and the second term is $o_\Pb(1/\sqrt{n})$ by Lemmas \ref{lemma:double_robustness} and \ref{lemma:kennedy}. We can therefore write 
\begin{align*}
     \riskhat(\Rt) - \risk(\Rt) &= (\Pn - \Pb)f_\theta + o_\Pb(1/\sqrt{n})
\end{align*}
By equivalent reasoning,
\begin{align*}
    \riskchangehat(\theta, \Rin) - \riskchange(\theta, \Rin) &= (\Pn - \Pb)(f_\theta - f_\thetat) + o_\Pb(1/\sqrt{n})
\end{align*}
Therefore, by the central limit theorem,
\begin{align*}
    & \sqrt{n}\left(\widehat{\risk}(\Rt) - \risk(\Rt)\right) \rightsquigarrow N\left(0, \var\left(f_\theta\right)\right) \\
    & \sqrt{n}\left(\riskchangehat - \riskchange\right) \rightsquigarrow N\left(0, \var\left(f_\theta - f_{\thetat}\right)\right)
\end{align*}
as claimed.

The reasoning for the fairness estimators is virtually identical. Let $\gamma_a = (1 - \phi)\one\{A = a\}$ and $\gammahat_a = (1 - \phihat)\one\{A = a\}$ as in the proof of Theorem \ref{thm:double_robustness}. From equation \eqref{eq:unfairness_expansion_final}, Proposition \ref{proposition:identification_derived}, and assumption A6 (nuisance estimator rates), we have
\begin{align*}
    &\widehat{\cFPR}(\Rt, a) - \cFPR(\Rt, a) = \\ 
    & \-\ \Pn[\gammahat_a]^{-1}(\Pn - \Pb)\Big\{(\theta_{a, 1} - \theta_{a, 0})\big(\Rin - \cFPR(\Rin, a)\big)\gamma_a\Big\} + o_\Pb(1/\sqrt{n})
\end{align*}
By the central limit theorem and Slutsky's theorem,
\begin{align*}
    \sqrt{n}\left(\widehat{\cFPR}(\Rt, a) - \cFPR(\Rt, a)\right) \rightsquigarrow N\left(0, \var\left(g_a\right)\right) 
\end{align*}
as claimed. The remaining statements follow by equivalent reasoning.

\section{Sample splitting} \label{appendix:sample_splitting}
A training sample, $\datatrain$, is used to estimate $\thetahat$, while a separate sample $\datatest$ is used to estimate the risk and fairness properties of the derived predictor $\Rhat$ conditional on $\thetahat$. Within each sample, separate folds are used to estimate the nuisance parameters $\mu_0$ and $\pihat$ versus the target parameters. 

The following schematic illustrates this procedure. $k$-fold cross fitting can be used within each sample to recover full sample size efficiency. For convenience, we suppose that each of the four samples is of size $n$, though our results require only that each sample is $O(n)$.
\begin{table}[h!]
    \centering
    \renewcommand{\arraystretch}{1.5}
    \begin{tabular}{ | P{1in} | P{1in} | }
         \multicolumn{2}{c}{\text{\large $\datatrain$}} \\
         \noalign{\vspace{2pt}}
         \hline
        $\datatrain^{\text{nuis}}$ & $\datatrain^{\text{target}}$ \\
        \hline
        \noalign{\vspace{-12pt}}
        \multicolumn{1}{@{}c@{}}{$\underbrace{\hspace*{\dimexpr\tabcolsep+2\arrayrulewidth}\hphantom{xxxxxxxxxx}}_{\substack{\\ \text{\large$\muhat_0, \-\ \pihat$}}}$}
        &
        \multicolumn{1}{@{}c@{}}{$\underbrace{\hspace*{\dimexpr\tabcolsep+2\arrayrulewidth}\hphantom{xxxxxxxxxx}}_{\substack{\\ \text{\large$\thetahat$}}}$}
    \end{tabular}
    \renewcommand{\arraystretch}{1.5}
    \begin{tabular}{ | P{1in} | P{1in} | }
         \multicolumn{2}{c}{\text{\large $\datatest$}} \\
        \noalign{\vspace{1.5pt}}
         \hline
        $\datatest^{\text{nuis}}$ & $\datatest^{\text{target}}$ \\
        \hline
        \noalign{\vspace{-12pt}}
        \multicolumn{1}{@{}c@{}}{$\underbrace{\hspace*{\dimexpr\tabcolsep+2\arrayrulewidth}\hphantom{xxxxxxxxxx}}_{\substack{\\ \text{\large$\muhat_0, \-\ \pihat$}}}$} &  
        \multicolumn{1}{@{}c@{}}{$\underbrace{\hspace*{\dimexpr\tabcolsep+2\arrayrulewidth}\hphantom{xxxxxxxxxx}}_{\substack{\\ \text{\normalsize{Properties of } \large{$\Rhat$}}}}$}
    \end{tabular}
    \label{t:sample_splitting}
\end{table}

\section{Simulations: data generating process} \label{appendix:simulations}
The data generating process used in section \ref{subsubsection:simulations1} to illustrate Theorems 1 and 2 is as follows, for data $Z = (A, X, \Rin, D, Y^0, Y^1, Y)$.
\begin{align*}
    \Pb(A = 1) &= 0.3 \\
    X \mid A &\sim \text{N}(A*(1, -0.8, 4, 2)^T, I_4) \\
    \Pb_{\text{pre}}(D = 1 \mid A, X) &= \min\{0.975, \expit((A, X)^T(0.2, -1, 1, -1, 1))\} \\
    \Pb_{\text{post}}(D = 1 \mid A, X, \Rin) &= \min\{0.975, \expit((A, X, \Rin)^T(0.2, -1, 1, -1, 1, 1))\} \\
    \Pb(Y^0 = 1 \mid A, X) &= \expit((A, X)^T(-5, 2, -3, 4, -5)) \\
    \Pb(Y^1 = 1 \mid A, X) &= \expit((A, X)^T(1, -2, 3, -4, 5)) \\
    Y &= (1 - D)Y^0 + DY^1
\end{align*}
where $I_4$ denotes the $4\times 4$ identity matrix and N denotes a Gaussian distribution. The predictor $\Rin(A, X)$ is trained using random forests. The pre-RAI decision making process doesn't depend on $\Rin$; the post-RAI process does.

For the simulations used in section \ref{subsubsection:simulations2} to illustrate fairness-performance tradeoffs, the distribution is identical except that $\Pb(Y^0 = 1 \mid A, X) = \expit((A, X)^T(-4, 0.4, 0.6, 0.8, -1))$.

\section{Asymptotic normality of doubly robust estimators} \label{appendix:asymptotic_normality}

To illustrate Theorem \ref{thm:asymptotic_normality}, an additional set of simulations was run using the post-RAI data generating process described above. First, $\theta$ was randomly set to $(0.74, 1.0, 0, 0.8)$. (Note that solutions to a linear program with a compact feasible set must occur at an extreme point of the set, so the presence of 0 and/or 1 in $\thetahat$ and $\thetaopt$ is virtually guaranteed.) The ``true'' risk $\risk(\Rt)$, risk change $\riskchange(\Rt)$, and error rate differences $\Deltapos(\Rt), \Deltaneg(\Rt)$ were again computed on a separate validation set of size 500,000, using plugin estimators with the true $\mu_0$. For conciseness, we omit results for $\widehat{\cFPR}$ and $\widehat{\cFNR}$. 

\begin{figure}[h]
    \centering
    \includegraphics[width=\linewidth]{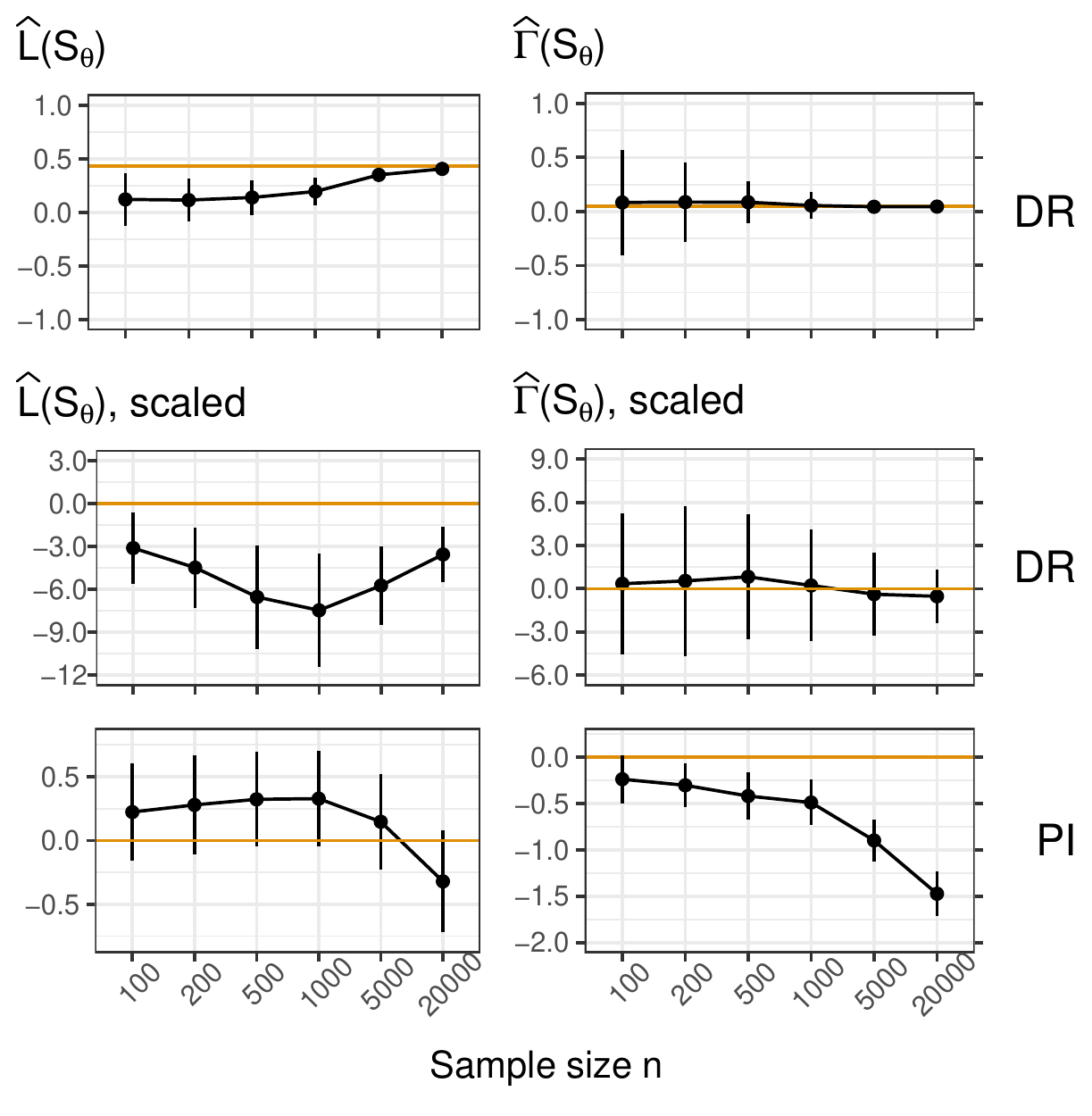}
    \caption{Doubly robust (DR) vs. plugin (PI) estimates of the loss and loss change for an arbitrary derived predictor $\Rt$, with $\theta = (0.74, 1.0, 0, 0.8)$, for samples of size 100 to 20,000.}
    \label{fig:simulations_risk_estimators}
\end{figure}

\begin{figure}[h]
    \centering
    \includegraphics[width=\linewidth]{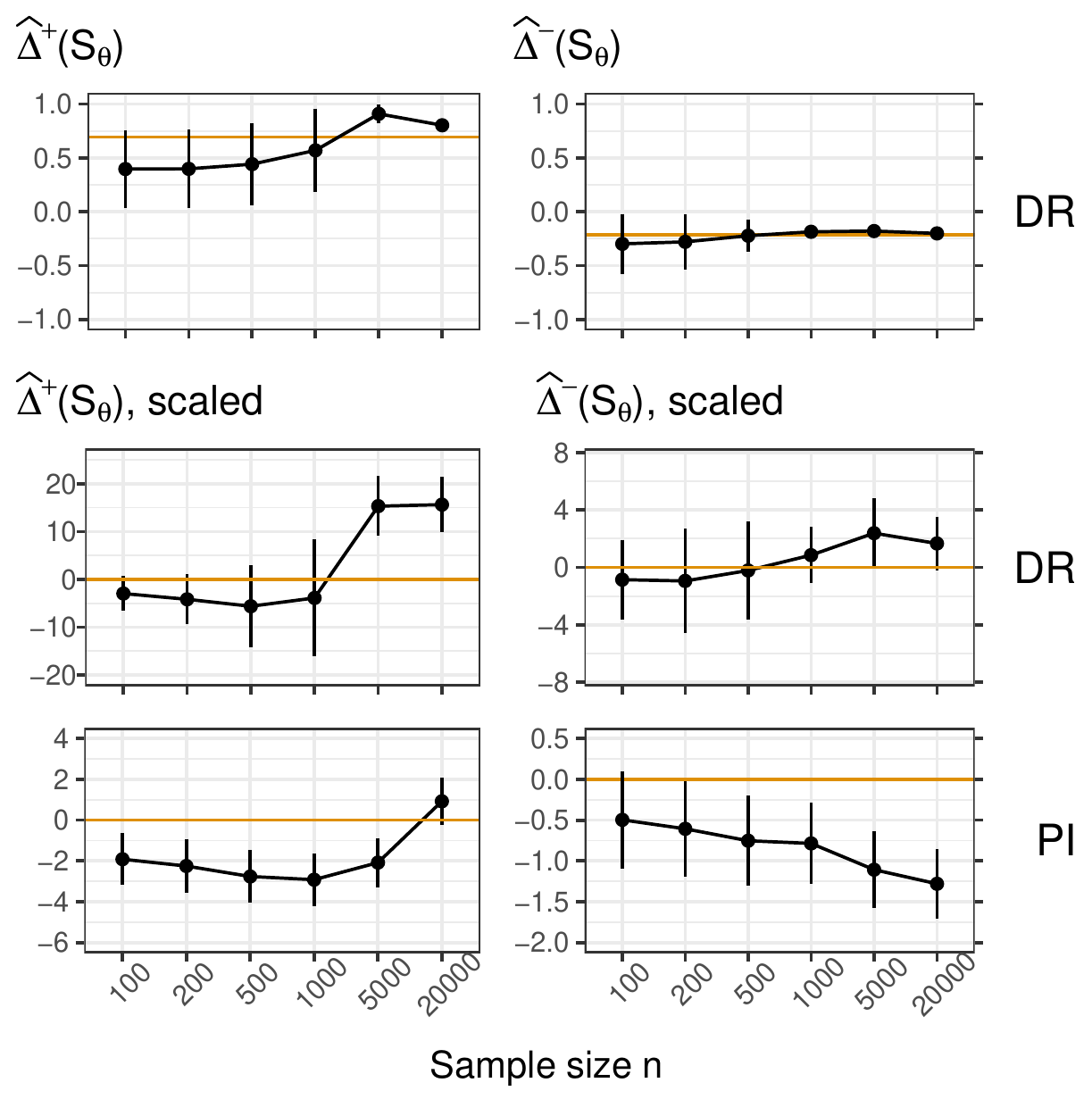}
    \caption{Doubly robust (DR) vs. plugin (PI) estimates of the error rate differences for an arbitrary derived predictor $\Rt$, with $\theta = (0.74, 1.0, 0, 0.8)$, for samples of size 100 to 20,000.}
    \label{fig:simulations_fairness_estimators}
\end{figure}

Figures \ref{fig:simulations_risk_estimators} and \ref{fig:simulations_fairness_estimators} illustrate results for doubly robust (DR) vs. plugin (PI) estimators of these quantities, for samples of size 100 to 20,000. Each vertical line represents a mean $\pm 1$ sd over 500 simulations. Orange horizontal lines represent the true parameter values (top rows in each figure) or 0. The top row shows that the doubly robust estimators converge to their target values. The bottom two rows represent values from the doubly robust and plugin estimators, transformed by $\psi(\Rhat) \mapsto \sqrt{n}(\psihat - \psi)$, where $\psihat, \psi$ are the relevant estimator and parameter for that column. These rows illustrate that $\sqrt{n}$-convergence is only guaranteed for the doubly robust estimators: the scaled values for the doubly robust estimators do not grow in $n$, while the scaled values for the plugin estimators begin to diverge (at least for $\riskchangehat$ and $\Deltahatneg$).

Table \ref{t:coverage} contains coverage results of 95\% confidence intervals for the error rates, error rate differences, loss, and loss change for the same arbitrary $\Rt$. The CIs were constructed using sample variances. To ensure that they did not exceed the bounds of the possible parameter values (i.e. $[0, 1]$ for the loss and error rates, $[-1, 1]$ for the error rate differences and loss change), the CIs were constructed using the delta method, via the transformations $\psihat \mapsto \logit(\psihat)$, for $\psihat \in \{\widehat{\cFPR}, \widehat{\cFNR}, \riskhat\}$, or $\psihat \mapsto \logit((\psihat + 1)/2)$, for $\psihat \in \{\Deltahatpos, \Deltahatneg, \riskchangehat\}$. Nominal coverage is achieved for various quantities at various sample sizes, but since the coverage guarantees are asymptotic, it is not surprising that it is not achieved everywhere. Interestingly, the median coverage rate in the table is 0.95. 

A separate set of CIs was computed without using the delta method; those results did not differ substantially and are therefore omitted here.
    
\begin{table}[ht]
\centering
    \renewcommand*{\arraystretch}{1.3}
\begin{tabular}{lrrrrrr}
\toprule
 &  100   &  200   &  500   &  1000  &  5000  &  20000 \\
\midrule
$\riskhat(\Rt)$        &   0.98 &   0.92 &   0.87 &   0.84 &   0.84 &   0.85 \\
\hline
$\riskchangehat(\Rt)$ &   1.00 &   0.99 &   0.93 &   0.94 &   0.71 &   0.78 \\
\hline
$\widehat{\cFPR}(\Rt, 0)$ &   0.99 &   0.98 &   0.98 &   0.96 &   0.94 &   0.95 \\
$\widehat{\cFPR}(\Rt, 1)$ &   0.90 &   0.89 &   0.93 &   0.95 &   0.96 &   0.57 \\
$\widehat{\cFNR}(\Rt, 0)$ &   0.99 &   0.99 &   0.98 &   0.99 &   0.92 &   0.93 \\
$\widehat{\cFNR}(\Rt, 1)$ &   0.99 &   0.99 &   0.99 &   1.00 &   0.98 &   0.71 \\
\hline
$\Deltahatpos(\Rt)$     &   0.98 &   0.98 &   0.97 &   0.99 &   0.97 &   0.92 \\
$\Deltahatneg(\Rt)$     &   0.99 &   1.00 &   0.99 &   0.99 &   0.94 &   0.94 \\
\bottomrule
\end{tabular}
    \caption{$95\%$ CI coverage at sample sizes ranging from $100$ to $20,000$ for the loss, loss change, error rates, and error rate differences, for an arbitrary derived predictor $\Rt$ with parameter $\theta = (0.74, 1.0, 0, 0.8)$. Coverage varies by estimator and sample size, though the median coverage is $95\%$.}
    \label{t:coverage}
\end{table}

\FloatBarrier
\onecolumn
\section{Notation} \label{appendix:notation}

{\renewcommand{\arraystretch}{1.3}
\begin{table*}[!h]
    \centering
    \label{t:notation2}
    \begin{tabular}{ | p{0.6\textwidth} | p{0.4\textwidth} | }
        \hline
        \multicolumn{2}{ | l | }{\textbf{Input data}} \\
        \hline
        $Z = (A, X, D, \Rin, Y) \sim \Pb$ & Sensitive feature $A$, covariates $X$, decision (treatment, intervention) $D$, input predictor $S$, outcome $Y$  \\
        \hline
        \multicolumn{2}{ | l | }{\textbf{Derived predictor}} \\
        \hline
        $\Rt \sim \Bern(\theta_{A,S})$ & Predictor derived from $S$ \\
        \hspace{1em} $\theta_{a, \rin} = \Pb(\Rt = 1 \mid A = a, \Rin = \rin)$ & Conditional probability that defines $\Rt$ \\ 
        \hspace{1em} $\theta_{A,\Rin} = \sum_{a, \rin\in \{0, 1\}} \one\{A=a, \Rin=\rin\}\theta_{a,\rin}$ & RV that takes value $\theta_{a,\rin}$ with probability $\Pb(A=a, \Rin=\rin)$ \\
        \hspace{1em} $\theta = (\theta_{0,0}, \theta_{0,1}, \theta_{1,0}, \theta_{1,1})^T$ & Optimization parameter \\
        \hspace{1em} $\thetat = (0, 1, 0, 1)$ & The value such that $\Rin_{\thetat} = \Rin$ \\
        \hline
        \multicolumn{2}{ | l | }{\textbf{Nuisance parameters}} \\
        \hline
        $\pi = \pi(A,X,\Rin) = \Pb(D = 1 \mid A,X,\Rin)$ & Propensity score for the decision \\
        $\mu_0 = \mu_0(A, X, \Rin, D) = \E[Y \mid A,X,\Rin,D=0]$ & Outcome regression \\
        $\phi = \frac{1-D}{1-\pi}(Y-\mu_0) + \mu_0$ & Uncentered influence function for $E[Y^0]$ \\
        \hline
        \multicolumn{2}{ | l | }{\textbf{Loss parameters}} \\
        \hline
        $\wfp, \wfn$ & Weights on the false positive and false negative rates \\
        $\beta_{a,\rin} = \E[\one\{A=a, \Rin=\rin\}(\wfp - (\wfp + \wfn)\mu_0)]$ & A coefficient in the loss, for $a, \rin \in \{0, 1\}$ \\
        $\beta = (\beta_{0,0}, \beta_{0,1}, \beta_{1,0}, \beta_{1,1})^T$ & Vector of loss coefficients \\
        $\risk(\Rt) = \wfp\Pb(\Rt = 1, Y^0 = 0) + \wfn\Pb(\Rt = 0, Y^0 = 1)$ & Loss of $\Rt$, equivalent to $\theta^T\beta + \wfn\E[\mu_0]$ \\
        $\riskchange(\Rt)$ & loss change $\risk(\Rt) - \risk(\Rin)$ \\
        \hline
        \multicolumn{2}{ | l | }{\textbf{Fairness parameters}} \\
        \hline
        $\cFPR(\Rt, a) = \Pb(\Rt = 1 \mid Y^0 = 0, A=a)$ & Counterfactual FPR for $\Rt$ for group $a$ \\
        $\cFNR(\Rt, a) = \Pb(\Rt = 0 \mid Y^0 = 1, A=a)$ & Counterfactual FNR for $\Rt$ for group $a$ \\
        $\betapos = (1 - \cFPR(\Rin, 0), \-\ \cFPR(\Rin, 0), \cFPR(\Rin, 1) - 1,\-\  -\cFPR(\Rin, 1))$ & \multirow{2}{*}{Coefficients for the fairness constraints} \\
        $\betaneg = (-\cFNR(\Rin, 0), \-\ \cFNR(\Rin, 0) - 1, cFNR(\Rin, 1), \-\ 1 - \cFNR(\Rin, 1))$ & \\
        $\Deltapos(\Rt) = \theta^T\betapos = \cFPR(\Rt, 0) - \cFPR(\Rt, 1)$ & Error rate differences of the predictor $\Rt$ in the $\cFPR$ \\
        $\Deltaneg(\Rt) = \theta^T\betaneg = \cFNR(\Rt, 0) - \cFNR(\Rt)$ & Error rate differences of the predictor $\Rt$ in the $\cFNR$ \\
        $\epsilonpos, \epsilonneg$ & Fairness constraints on $\Deltapos$ and $\Deltaneg$ \\
        $\UFP(\Rt) = \max( \mid \Deltapos(\Rt) \mid  - \-\ \epsilonpos, 0)$ & Excess unfairness in the $\cFPR$ \\
        $\UFN(\Rt) = \max( \mid \Deltaneg(\Rt) \mid  - \-\ \epsilonneg, 0)$ & Excess unfairness in the $\cFNR$ \\
        \hline
        \multicolumn{2}{ | l | }{\textbf{Optimal fair derived predictor}} \\
        \hline
        $\argmin_{\theta}\risk(\Rt) \text{ s.t. } | \Deltapos(\Rt) |  \leq \epsilonpos,  | \Deltaneg(\Rt) |  \leq \epsilonneg$ & Parameter defining the optimal fair derived predictor $\Ropt$ \\
        \hline
    \end{tabular}
\end{table*}
}

\end{document}